\documentclass[twocolumn,nofootinbib,aps,showpacs,floats,letterpaper,floatfix,groupedaddress]{revtex4}

\usepackage{graphicx,epsf, epsfig, amssymb}% Include figure files
\usepackage{amsfonts,amsmath,amssymb,amsthm}
\usepackage{bm}
\usepackage{dcolumn}
\usepackage[latin1]{inputenc}
\usepackage{latexsym}
\usepackage{rotating}
\usepackage{hyperref}
\usepackage[dvips]{color}
\usepackage{longtable}
\usepackage{enumerate}

\def\be{\begin{equation}}
\def\ee{\end{equation}}
\def\beq{\begin{eqnarray}}
\def\eeq{\end{eqnarray}}

\def\nn{\nonumber}

\begin{document}

\title{Semianalytical estimates of scattering thresholds and gravitational
  radiation\\in ultrarelativistic black hole encounters}

\author{Emanuele Berti$^{1,2}\footnote{Electronic address: berti@phy.olemiss.edu}$,
Vitor Cardoso$^{1,3}$,
%\footnote{Electronic address: vitor.cardoso@ist.utl.pt}$,
%
Tanja Hinderer$^{2}$,
%\footnote{Electronic address: tph25@caltech.edu}$,
%
Madalena Lemos$^{3}$,
%\footnote{Electronic address: madalena.dal@gmail.com}$,
%
Frans Pretorius$^{4}$,
%\footnote{Electronic address: fpretori@princeton.edu}$,
%
Ulrich Sperhake$^{1,2,5}$,
%\footnote{Electronic address: sperhake@tapir.caltech.edu}$,
%
Nicol\'{a}s Yunes$^{4}$%\footnote{Electronic address: nyunes@princeton.edu}$
}
\affiliation{${^1}$ Department of Physics and Astronomy, The University of
Mississippi, University, MS 38677, USA}
\affiliation{${^2}$ California Institute of Technology, Pasadena, CA 91109, USA}

\affiliation{${^3}$ CENTRA,~Dept.~de F\'{\i}sica,~Instituto~Superior~T\'ecnico, Av.~Rovisco Pais 1, 1049 Lisboa, Portugal}
\affiliation{${^4}$ Department of Physics, Princeton University, Princeton, NJ 08544, USA}
\affiliation{${^5}$ Institut de Ci\`encies de l'Espai (CSIC-IEEC), Facultat de
  Ci\`encies, Campus UAB, Torre C5 parells, Bellaterra, 08193, Spain}

%\date{\today}

\begin{abstract}
Ultrarelativistic collisions of black holes are ideal gedanken experiments to
study the nonlinearities of general relativity. In this paper we use
semianalytical tools to better understand the nature of these collisions and
the emitted gravitational radiation. We explain many features of the energy
spectra extracted from numerical relativity simulations using two
complementary semianalytical calculations. In the first calculation we
estimate the radiation by a ``zero-frequency limit'' analysis of the collision
of two point particles with finite impact parameter. In the second calculation
we replace one of the black holes by a point particle plunging with arbitrary
energy and impact parameter into a Schwarzschild black hole, and we explore
the multipolar structure of the radiation paying particular attention to the
near-critical regime.  We also use a geodesic analogy to provide qualitative
estimates of the dependence of the scattering threshold on the black hole spin
and on the dimensionality of the spacetime.
\end{abstract}

%\begin{widetext}
%\tableofcontents
%\end{widetext}
%\clearpage


%04.25.D-    Numerical relativity
%04.25.dc    Numerical studies of critical behavior, singularities, and cosmic censorship
%04.25.dg    Numerical studies of black holes and black-hole binaries
%04.25.-g    general relativity: approximation methods, equations of motion
%04.50.-h    Higher-dimensional gravity and other theories of gravity
%04.50.Cd    KaluzaKlein theories
%04.50.Gh    Higher-dimensional black holes, black strings, and related objects
%04.60.Cf    Gravitational aspects of string theory
%04.70.-s    Physics of black holes
%04.70.Bw    Classical black holes
%04.70.Dy    Quantum aspects of black holes, evaporation, thermodynamics
%04.80.-y    Experimental studies of gravity
%04.80.Cc    Experimental tests of gravitational theories
%11.25.Mj    Compactification and four-dimensional models
%11.10.Kk    Field theories in dimensions other than four

\maketitle

%\tableofcontents

\newpage
%%%%%%%%%%%%%%%%%%%%%%%%%%%%%%%%%%%%%%%%%%%%%%%%%%%%%%%%%%%%%%%%%%%%%%%%%%%%%%%%
\section{Introduction}
%%%%%%%%%%%%%%%%%%%%%%%%%%%%%%%%%%%%%%%%%%%%%%%%%%%%%%%%%%%%%%%%%%%%%%%%%%%%%%%%

Solving Einstein's equations numerically is a highly nontrivial task. After
the recent breakthroughs in numerical relativity (NR)
\cite{Pretorius:2005gq,Baker:2005vv,Campanelli:2005dd}, binary black hole (BH)
mergers have been routinely carried out by several groups worldwide. As is
often the case in physics, the results of numerical simulations are best
understood or interpreted by studying simplified models that capture the main
features of the problem.  In this paper we will argue that semianalytical
tools are particularly useful to gain insight into the fascinating but complex
problem of ultrarelativistic BH collisions.

High-energy BH encounters are a difficult undertaking in NR (see
Refs.~\cite{Sperhake:2008ga,Shibata:2008rq,Sperhake:2009jz,Choptuik:2009ww}
for a discussion of ultrarelativistic collisions in four dimensions,
Refs.~\cite{Yoshino:2009xp,Nakao:2009dc,Shibata:2009ad,Zilhao:2010sr} for
recent progress on $D$-dimensional simulations and Ref.~\cite{Cardoso:2005jq}
for a review of the most outstanding questions in ultrarelativistic BH
collisions {\rm before} the recent breakthroughs in NR). Efficient adaptive
mesh refinement and wave extraction techniques are required because the
problem involves various scales: the BHs are ``pancake-shaped'' because of
Lorentz contraction, the large speeds involved in the collision require large
initial separations to define asymptotic states, and high resolution is
required to study the dynamics of the final BH (if any) formed as a result of
the merger. Further difficulties arise from the spurious radiation present in
the initial data.

Despite their relatively limited accuracy, simulations of ultrarelativistic BH
collisions to date have provided definite predictions for nonspinning BHs in four dimensions
\cite{Sperhake:2008ga,Shibata:2008rq,Sperhake:2009jz,Choptuik:2009ww}. For
example these simulations show that, even in the highly symmetric case of
head-on collisions, as much as $\sim 14\pm 3$\% of the energy of the system
can be radiated in gravitational waves \cite{Sperhake:2008ga,Shibata:2008rq}.
They further demonstrate the existence of three distinct regimes depending on
the impact parameter $b$: immediate mergers, nonprompt mergers and the
scattering regime.  These regimes are separated by two special values of the
impact parameter: the {\em threshold of immediate merger} $b^*$ and the {\em
  scattering threshold} $b_{\rm scat}$. Roughly speaking, for $b < b^*$ merger
occurs within the first encounter, whereas for $b^* < b < b_{\rm scat}$ it
does not, but sufficient energy is radiated to put the binary into a bound
state that eventually results in a merger. For the largest initial center of
mass velocities of $v=0.94$ studied to date, as much as $\sim 35\pm 5$\% of
the energy can be carried away by gravitational radiation
\cite{Sperhake:2009jz,Shibata:2008rq}.

Close to the threshold of immediate merger, the binary exhibits ``zoom-whirl''
behavior, an extreme version of relativistic perihelion precession which can
be precisely defined and understood in the geodesic limit
\cite{Merrick:2007,Pretorius:2007jn,Grossman:2008yk,Healy:2009zm,Gold:2009hr}.
For point particles orbiting BHs, zoom-whirl orbits are intimately related to
the existence of unstable spherical orbits (unstable {\em circular} orbits at
radii $3M \leq r \leq 6M$ in the special case of Schwarzschild BHs).  The very
existence of zoom-whirl orbits in comparable-mass BH encounters is perhaps a
strong hint that simple, semianalytical approaches (in this case, the study of
point-particle geodesics around BHs) can provide valuable insight into the
general solution of the problem.

\begin{figure}[htb]
\begin{center}
\includegraphics[scale=0.33,angle=270,clip=true]{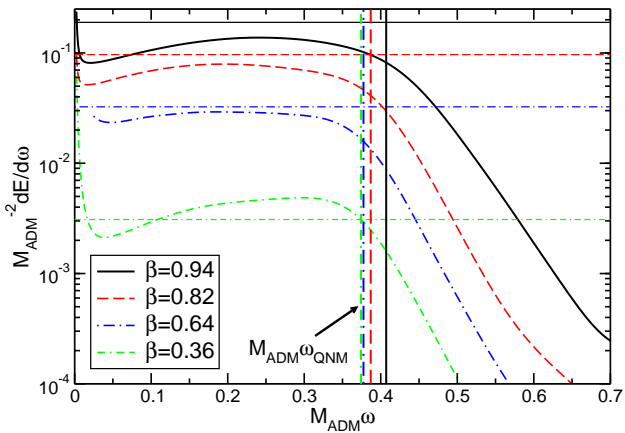}
\end{center}
\caption{\label{fig:headonspec} Energy spectrum for the dominant (quadrupolar,
  i.e. $l=2$) component of the gravitational radiation computed from NR
  simulations of the head-on collision of two equal-mass BHs (from
  \cite{Sperhake:2008ga}). The collision speed in the center-of-mass frame,
  $\beta=v/c$, is indicated in the legend. The energy spectrum is roughly flat
  (independent of frequency) up to the quasinormal mode (QNM) frequencies
  (marked by vertical lines), after which it decays exponentially. All
  quantities are normalized to the Arnowitt-Deser-Misner (ADM) mass of the
  system $M_{\rm ADM}$.}
\end{figure}

NR simulations of ultrarelativistic BH collisions have also provided
information on the structure of the Fourier-domain energy spectra $dE^{\rm
  rad}/d\omega$.  An interesting feature of the spectra of comparable-mass BH
collisions is the presence of a nonvanishing zero-frequency limit (ZFL). For
head-on collisions, Fig.~\ref{fig:headonspec} shows that the energy spectrum
is roughly flat (independent of frequency) up to some cutoff frequency, after
which it decays exponentially (in this figure and elsewhere in the paper we
use geometrical units $G=c=1$). Grazing ultrarelativistic BH collisions also
lead to energy spectra with complex features, although this was not discussed
in Ref.~\cite{Sperhake:2009jz} because of space limitations.  These complex
spectra are discussed further in section \ref{sec:comp} of the present work.

One of the motivations of this paper is to explore simplified, analytical and
semianalytical descriptions of ultrarelativistic BH mergers that provide
qualitative (and sometimes even quantitative) explanations of the main
features of these spectra \cite{Sperhake:2008ga}. In particular, we shall
concentrate on the combination of the following two approaches: a linearized
calculation in flat spacetime pioneered by Weinberg and Smarr
\cite{Weinberg:1964ew,Weinberg:1965nx,Smarr:1977fy,Adler:1975dj} and BH
perturbation theory
\cite{Ruffini:1973ky,Ferrari:1981dh,Lousto:1996sx,Cardoso:2002ay,Cardoso:2002yj,Cardoso:2002jr,Berti:2003si}.
We will refer to the Weinberg-Smarr approach (which is valid for arbitrary
velocities, as long as the radiated energies are small) as the ``ZFL
calculation,'' because it provides a good approximation of the emitted energy
spectrum at low frequencies. The ZFL calculation assumes an instantaneous
collision in flat spacetime, and in four dimensions it yields a flat spectrum
{\it at all frequencies} -- precisely what one would expect from the analogous
problem in electromagnetism \cite{jackson}. The origin of the exponential
cutoff in the NR spectra can be understood through the second approach,
i.e. by computing the radiation from point particles falling in a {\it curved}
BH spacetime
\cite{Ruffini:1973ky,Ferrari:1981dh,Lousto:1996sx,Cardoso:2002ay,Cardoso:2002yj,Cardoso:2002jr,Berti:2003si}.
These perturbative calculations show that the radiation in a given multipole
``shuts off'' at the real part of the lowest QNM frequency of the BH for the
multipole in question \cite{Berti:2005ys,Berti:2009kk}.

In this paper we address several questions related to the puzzling behavior of
ultrarelativistic BH collisions:
\begin{enumerate}
\item We use geodesic calculations to explore the dependence of the scattering
  threshold on BH spin and dimensionality of the spacetime. This threshold
  determines whether the two BHs will eventually merge or not, and it is of
  fundamental importance to estimate cross sections for BH production in
  high-energy collisions \cite{Kanti:2004nr}. Classical general relativity in
  $D$ dimensions should be adequate to determine the cross section for BH
  production at trans-Planckian collision energies and the fractions of the
  collision energy and angular momentum lost in gravitational radiation. This
  information will be of paramount importance to improve the modeling of
  microscopic BH production in event generators such as {\sc Truenoir}, {\sc
    Charybdis2}, {\sc Catfish} or {\sc Blackmax}
  \cite{Dimopoulos:2001hw,Frost:2009cf,Cavaglia:2006uk,Dai:2009by}. The event
  generators will then provide a description of the corresponding evaporation
  phase, which might be observed during LHC collisions.

\item We reproduce the qualitative (and sometimes quantitative) features of
  the energy spectra by a ``bare-bones'' approach, where we replace the
  colliding BHs by two point particles and we estimate the radiation by a ZFL
  calculation with {\it finite} impact parameter. The generalization to finite
  impact parameter requires a specific model to bind the particles after the
  collision and guarantee energy-momentum conservation \cite{Price:1973ns},
  but our qualitative conclusions should be roughly independent of the details
  of the model.

\item We gain insight into the details of the emitted radiation (particularly
  in the near-critical regime) by a perturbative study of the
  ultrarelativistic infall of point particles with generic impact parameter
  into a Schwarzschild BH.
\end{enumerate}

The two-pronged analytical approach above provides a better
understanding of most of the main features of the energy spectra. In
particular, we reach the following conclusions:
\begin{enumerate}[i.]
\item In both the extreme-mass ratio case and in comparable-mass NR
  simulations, the ZFL is by and large independent of the impact parameter
  $b$.
\item For a given multipolar index $l$, the slope of the different
  $m$-components at low frequencies is proportional to $b$.
\item The exponential cutoff is related to the QNM frequencies of the final
  BH.
\item Higher multipoles of the radiation become increasingly important as the
  binary becomes ultrarelativistic, leading from an exponential scaling of the
  form $e^{-a_R \;l}$ in the nonrelativistic limit to a scaling of the form
  $b_R/l$ in the ultrarelativistic limit, where $(a_R,b_R)$ are constants.
\item The total radiation is proportional to the number of orbits near
  the scattering threshold for {\it nonrelativistic} infalls, but it is
  enormously enhanced (essentially by a resonance with the QNM
  frequencies of the final BH) when the collision is ultrarelativistic.
\end{enumerate}

The plan of the paper is as follows. In section~\ref{sec:criticalb} we compute
the critical impact parameter for geodesics in four and higher dimensions.
In section~\ref{zflgrazing} we explore a simple toy model to estimate the
radiation in the ZFL for collisions with finite impact parameter. In
section~\ref{ppart} we compute the energy, angular momentum and linear
momentum radiated by point particles falling into Schwarzschild BHs in four
dimensions with arbitrary energy. In section~\ref{sec:comp} we show how the
ZFL and point-particle calculations shed light on some features of NR
simulations of ultrarelativistic BH collisions. We conclude by pointing out
possible extensions of our investigation. Appendix \ref{sec:SNformalism} gives
some technical details on the Sasaki-Nakamura formalism, and Appendix
\ref{app:multipoles} shows how the ZFL results can be decomposed in terms of
spin-weighted spherical harmonics.

We work mostly in four spacetime dimensions with signature $(-,+,+,+)$
\cite{Misner:1973cw}, except for the discussion in section~\ref{MPBHs}, where
we explore higher-dimensional BHs. Greek letters $(\mu,\nu,\ldots)$ in index
lists range over all spacetime indices. The
Einstein summation convention is employed unless otherwise specified.
Overhead dots stand for partial
differentiation with respect to the proper time $\tau$, $\dot{E} \equiv
\partial E/\partial \tau$; $\mathcal{O}(A)$ stands for terms of order $A$.

%%%%%%%%%%%%%%%%%%%%%%%%%%%%%%%%%%%%%%%%%%%%%%%%%%%%%%%%%%%%%%%%%%%%%%%%%%%%%%%%
\section{Scattering threshold in black hole spacetimes\label{sec:criticalb}}
%%%%%%%%%%%%%%%%%%%%%%%%%%%%%%%%%%%%%%%%%%%%%%%%%%%%%%%%%%%%%%%%%%%%%%%%%%%%%%%%

One of the main ingredients to estimate BH production rates in particle
accelerators is the cross section for BH production
\cite{Kanti:2004nr}. Several analytical approximations have been used to
estimate cross sections \cite{Cardoso:2005jq}.  In this section we explore a
particularly simple approach. We compute the critical impact parameter between
plunge and scattering for particles following geodesics in different BH
metrics, in order to estimate the qualitative dependence of the cross section
on BH spin and dimensionality. We begin with a brief review of the scattering
threshold in the Schwarzschild background. Then we proceed to the computation
of the critical impact parameter in a four-dimensional rotating (Kerr)
background, and finally we generalize the analysis to higher-dimensional
rotating (Myers-Perry) backgrounds.

%%%%%%%%%%%%%%%%%%%%%%%%%%%%%%%%%%%%%%%%%%%%%%%%%%%%%%%%%%%%%%%%%%%%%%%%%%%%%%%%
\subsection{Schwarzschild black holes\label{sec:Lcrit}}
%%%%%%%%%%%%%%%%%%%%%%%%%%%%%%%%%%%%%%%%%%%%%%%%%%%%%%%%%%%%%%%%%%%%%%%%%%%%%%%%

Geodesics in a Schwarzschild background are completely determined by their
energy and ($z$-component of) orbital angular momentum per unit rest mass.  It
is useful to replace the total energy $\tilde E$ and the orbital angular
momentum $\tilde L_z$ by the parameters $E=\tilde E/\mu$ and $L_z=\tilde
L_z/\mu$. Here $\mu$ is the particle's rest mass and $E$ is related to the
velocity $v$ of the particle at infinity by $E = (1 - v^{2})^{-1/2}$.  The
geodesic radial behavior is governed by the relation $(d r/d\tau)^2=E^2-V_{\rm
  eff}(r,L_z)$, where $\tau$ is the proper time and the effective potential
for a Schwarzschild background is $V_{\rm eff} = (1+L_z^2/r^2) f$, with $f
\equiv 1 - 2 M/r$ and $M$ the BH mass.

Geodesics can be classified according to how their energy compares to the
maximum value of the effective potential, which in this case is given by
\be V^{\rm max}_{\rm eff}=
\frac{1}{54}\left[\frac{L_z^2}{M^2}+36+\left(\frac{L_z^2}{M^2}-12\right)
\sqrt{1-\frac{12}{(L_z^2/M^2)}}\right]. \label{max-Veff} \ee
The scattering threshold is then defined by the condition $E^2=V^{\rm
  max}_{\rm eff}$: unbound orbits with $E^2>V^{\rm max}_{\rm eff}$ are
captured, while those with $E^2<V^{\rm max}_{\rm eff}$ are scattered. Given
some $E$, the critical radius or impact parameter $b_{\rm crit}$ that defines
the scattering threshold is obtained by solving the condition $E^2=V^{\rm
  max}_{\rm eff}$ for $L_{z} = L_{\rm crit}$, and then using the relation
\be b_{\rm crit} = L_{\rm crit} (E^{2} - 1)^{-1/2}\,. \label{bcrit-eq}
\ee
Note that we mostly work in the point particle approximation with no radiation
reaction effects, so we adopt a slightly different terminology from that of
Ref.~\cite{Sperhake:2009jz}: here $L_{\rm crit}$ denotes the critical angular
momentum separating plunging trajectories from scattering trajectories.

Ultrarelativistic collisions can be modeled by large-energy geodesics. One can
then show that a high-energy orbit plunges when
\be \frac{L_z^2}{M^2}< \frac{L_{\rm crit}^2}{M^2}\simeq
\frac{27}{(1/E^2)}-9-\frac{1}{E^2}+{\cal O}(1/E^4)\,,
\label{urlzcrit} 
\ee
where the last relation is valid in the limit $E^{2} \gg 1$. The condition for
scattering is $L_z> L_{\rm crit}$. One can also easily show that all orbits
plunge if $L_z^2<12 M^2$ when the particle is at rest at infinity ($E=1$).
If $E>1$,
% then all orbits plunge if $12 M^2<L_z^2<16 M^2$, while
scattering orbits exist only if $L_{z} > 4 M$.

\subsection{Kerr black holes}
%%%%%%%%%%%%%%%%%%%%%%%%%%%%%%%%%%%%%%%%%%%%%%%%%%%%%%%%%%%%%%%%%%%%%%%%%%%

The radial motion of equatorial geodesics in a Kerr background can also be
described in terms of an effective potential \cite{Bardeen:1972fi}. In this
case, however, the potential for timelike geodesics takes a more complicated
form, which we can parametrize as
\be V_{\rm eff} =  \left(1 + \frac{L_{z}^{2}}{r^{2}} \right) f + j \;
\frac{\sigma_{1} L_{z}}{r^{3}} + j^{2} \; \left(
\frac{\sigma_{2}}{r^{2}} +  \frac{\sigma_{3}}{r^{3}} \right)\,. \ee
The Kerr spin parameter $a$ and the reduced spin parameter $j$ are related to
the spin angular momentum $J$ via $J = M a = j M^{2}$. As before, $E$ and
$L_{z}$ stand for the energy and ($z$-component of) angular momentum per unit
rest mass.  For convenience, we have also defined the energy-dependent
``spin-deformation'' parameters
\be 
\sigma_{1} \equiv 4 M^{2} \; E\,,\quad
\sigma_{2} \equiv M^{2} \left(1-E^{2}\right)\,,\quad
\sigma_{3} \equiv - 2 M^3 E^2\,. 
\ee

For fixed $j$, orbits with a turning point are identified by pairs $(E,L_{z})$
for which $V_{\rm eff}$ has a double root. The extrema of the effective
potential are defined by
\be \frac{d V_{\rm eff}}{dr} = \frac{2 M}{r^{2}} + \frac{6 M
L_{z}^{2}}{r^{4}} - \frac{2 L_{z}^{2}}{r^{3}} - \frac{3 j \sigma_{1}
L_{z}}{r^{4}} - j^{2} \left( \frac{2 \sigma_{2}}{r^{3}} + \frac{3
\sigma_{3}}{r^{4}} \right)=0 \ee
and correspond to radii
\beq r_{\pm} &=& \frac{L_{z}^{2}}{2M} \left(1 + j^{2}
\frac{\sigma_{2}}{L_{z}^{2}} \right) \pm
 \frac{L_{z}^{2}}{2 M}
 \left[1 - \frac{12 M^{2}}{L_{z}^{2}} + j \frac{6 M \sigma_{1}}{L_{z}^{3}}
 \right.
 \nonumber \\
 &+& \left.
  j^{2} \left( \frac{2 \sigma_{2}}{L_{z}^{2}}
 + \frac{6 M \sigma_{3}}{L_{z}^{4}} \right) + j^{4} \frac{\sigma_{2}^{2}}{L_{z}^{4}} \right]^{1/2}.
\eeq
The maximum of the effective potential is then equal to $V_{\rm eff}^{\rm max}
= V_{\rm eff}(r = r_{-})$, which generalizes Eq.~\eqref{max-Veff}.

\begin{figure}[htb]
\begin{center}
\includegraphics[scale=0.33,clip=true]{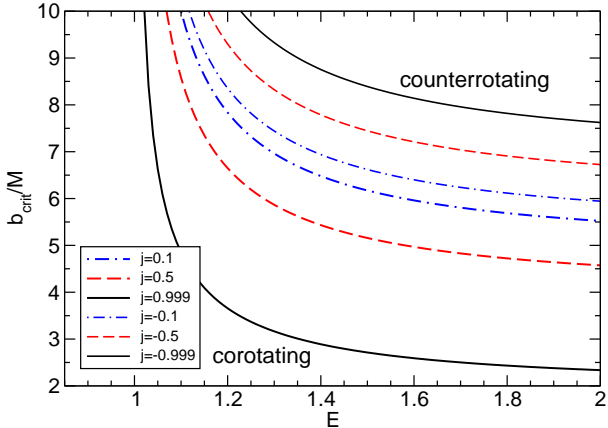}
\end{center}
\caption{\label{bccontour} Critical impact parameter versus reduced energy $E$
  for different values of $j$ in a Kerr background.}
\end{figure}

If we are given some values of $(E,j)$, the critical impact parameter $b_{\rm
  crit}$ can be obtained by solving $E^{2} = V_{\rm eff}^{\rm max}$ for
$L_{\rm crit}$, and then using Eq.~\eqref{bcrit-eq} .  Figure~\ref{bccontour}
plots this critical impact parameter as a function of energy for different
values of the spin, where we have solved for $b_{\rm crit}$ numerically.
Observe that $b_{\rm crit}$ asymptotes rapidly to a constant as the energy
approaches $E \sim 2$. For large, positive spins (co-rotating case), the
critical impact parameter asymptotes to $b_{\rm crit} \sim 2.33 M$ for $E =
2$, while for large, negative spins (counter-rotating case) it asymptotes to
$b_{\rm crit} \sim 7.62 M$.  As the spin is decreased, the critical impact
parameter for corotating geodesics increases, so that the smallest impact
parameter corresponds to maximally spinning BHs.

The above considerations suggest that an asymptotic analysis might allow for a
better analytical understanding of the critical impact parameter in the
ultrarelativistic regime.  Let us then define the perturbation parameters
\be 
1 \gg \epsilon \equiv 1 - j  > 0\,, \qquad 
1 \gg \xi \equiv 1/E>0\,. 
\ee
Performing an asymptotic expansion to second order in $\epsilon$, the solution
to $E^{2} = V_{\rm eff}^{\rm max}$ becomes
\beq L_{\rm crit} &\sim& \frac{2}{\xi} + \sqrt{\epsilon}
\frac{\sqrt{6 - 2 \xi^{2}}}{\xi} + \epsilon \frac{2 \left( \xi^{2} -
1\right)}{\xi \left(\xi^{3} - 3\right)}
\\ \nonumber
&+& \epsilon^{3/2} \frac{\left(\xi^{2} - 1 \right) \left(3 \xi^{4} +
5\right)
  \sqrt{6 - 2 \xi^{2}}}{\xi \left(\xi^{2} - 3 \right)^{3}}
+ {\cal{O}}(\epsilon^{2}). \eeq
Reexpanding this equation to second order in $\xi$, we find
\beq \label{asymptbc} 
\frac{b_{\rm crit}}{M} &\sim& 2 + \sqrt{6}
\sqrt{\epsilon} + \frac{2}{3} \epsilon  + \frac{5 \sqrt{6}}{108}
\epsilon^{3/2}
 \\ \nonumber
&+& \left(1 + \frac{\sqrt{6}}{3} \sqrt{\epsilon} - \frac{1}{9}
\epsilon + \frac{5 \sqrt{6}}{324} \epsilon^{3/2} \right) \xi^{2} +
{\cal{O}}(\epsilon^{2},\xi^3). \eeq
Eq.~\eqref{asymptbc} is formally a bivariate expansion, i.e. an expansion in
two independent perturbation parameters: $\epsilon$ and $\xi$.  This expansion
is a fractional Frobenius series, with the regular limit $b_{\rm crit} \to 2$
when $\epsilon \to 0$ and $\xi \to 0$.  The energy parameter contributes only
to second order, while the spin contributes at fractional leading order. A
comparison of this estimate to the numerical solution of Fig.~\ref{bccontour}
shows good agreement: when $j=0.999$ and $E = 2$ Eq.~\eqref{asymptbc}
predicts
%
%$b_{\rm crit} \sim 2.33455741299372 M$, 
%
$b_{\rm crit} \sim 2.335 M$, while the numerical result is
%
%$b_{\rm num} = 2.39566974783873 M$, 
$b_{\rm num} = 2.396 M$. Therefore the relative fractional error of the
asymptotic expansion (dominated by the neglected relative $E^{-3}$ terms) is
approximately $2.6\%$.

In the ultrarelativistic limit ($\xi \to 0$), we can solve the critical impact
parameter equation exactly for arbitrary $j$ to find \cite{Bardeen:1972fi}
\be \frac{b_{\rm crit}}{M}=-j +
6\cos\left[\frac{\arccos(-j)}{3}\right]\,, 
\label{4D-bcrit} 
\ee
or alternatively $b_{\rm crit}/M = - j + 3 x + 3/x$, where
$x \equiv [-j +(j^2-1)^{1/2}]^{1/3}$. The asymptotic expansion is very
accurate in the ultrarelativistic limit: for example, for $j =0.999$ this
exact formula gives $b_{\rm crit}/M = 2.07812987$, to be compared with $b_{\rm
  crit}/M \sim 2.07812992$ from Eq.~(\ref{asymptbc}).

%%%%%%%%%%%%%%%%%%%%%%%%%%%%%%%%%%%%%%%%%%%%%%%%%%%%%%%%%%%%%%%%
\subsection{Myers-Perry black holes\label{MPBHs}}
%%%%%%%%%%%%%%%%%%%%%%%%%%%%%%%%%%%%%%%%%%%%%%%%%%%%%%%%%%%%%%%%

We now generalize the above analysis to higher-dimensional BHs,
considering for illustration the Myers-Perry solution with a single angular
momentum direction \cite{Myers:1986un}. 
%%%%%%%%%%%%%%%%%%%%%%%%%%%%%%%%%%%%%%%%%%%%%%%%%%%%%%%%%%%%%%%%%%%%%%%%%%%%%
The metric of a $D$-dimensional Kerr BH with only one nonzero angular momentum
parameter is given in Boyer-Lindquist-type coordinates by \cite{Myers:1986un}
\begin{eqnarray}
ds^2&=&
-{\Delta-a^2\sin^2\vartheta\over\Sigma}dt^2
-{2a(r^2+a^2-\Delta)\sin^2\vartheta \over\Sigma}
dtd\varphi \nonumber\\
&&{}+{(r^2+a^2)^2-\Delta a^2 \sin^2\vartheta\over\Sigma}\sin^2\vartheta d\varphi^2
\nonumber\\
&&{}
+{\Sigma\over\Delta}dr^2
+{\Sigma}d\vartheta^2+r^2\cos^2\vartheta d\Omega_{D-4}^2,
\label{metric}
\end{eqnarray}
where
\begin{eqnarray}
\Sigma&=&r^2+a^2\cos^2\vartheta,\\
\Delta&=&r^2+a^2-M_* r^{5-D},
\end{eqnarray}
$d\Omega_{D-4}^2$ denotes the standard metric of the unit $(D-4)$-sphere, and
$M_{*}$ is related to the BH mass. In fact this metric describes a rotating BH
in an asymptotically flat vacuum spacetime with mass ${\cal M}$ and angular
momentum ${\cal J}$ given by
\be
 {\cal M}=\frac{D-2}{16\pi }A_{(D-2)}M_*,\quad
 {\cal J}=\frac{1}{8\pi }A_{(D-2)}M_* a\,,
\ee
where $A_{(D-2)}$ is the area of a unit ($D-2$)-sphere:
\beq
 A_{(D-2)}=\frac{2\pi ^{(D-1)/2}}{\Gamma [(D-1)/2]}\,.
\eeq
%
%%%%%%%%%%%%%%%%%%%%%%%%%%%%%%%%%%%%%%%%%%%%%%%%%%%%%%%%%%%%%%%%%%%%%%%%%%%%

Timelike equatorial geodesics depend on the effective potential
\be V_{\rm eff} = r^{2} E^{2} + \frac{{M_*}}{r^{D-3}} \left(a E -
L_{z}\right)^{2} + \left(a^{2} E^{2} - L_{z}^{2} \right) - \delta_{1}
\Delta_{D}, \ee
where $D$ is the dimensionality of spacetime and $\Delta_{D} = r^{2} + a^{2} -
{M_*} r^{5-D}$.  For $D=4$, the quantity ${M_*}$ and the mass of the BH are
related via ${M_*} = 2 {\cal M}$. In higher dimensions $M_*$ does not have
units of length, but instead $[M_*] = ({\rm length})^{D-3}$.  We also introduce
a dimensionless spin parameter
\be
j=\frac{a}{(M_*/2)^{1/(D-3)}}\,,
\ee
which reduces to the corresponding Kerr quantity for $D=4$.

When $D=5$ we can carry out an asymptotic analysis similar to the one
presented in the previous subsection (see also \cite{Cardoso:2008bp}). In the
ultrarelativistic limit ($E \to \infty$) we have
\be
b_{\rm crit}^{(D=5)}/\sqrt{M_*}=2 - j/\sqrt{2}\,, 
\ee
which is to be contrasted with the four-dimensional result presented in
Eq.~\eqref{4D-bcrit}.  This asymptotic analysis reveals that both in four and
five dimensions the overall effect of spin for corotating geodesics is to
reduce the critical impact parameter, suggesting that the highest energy
emission occurs for maximally rotating BHs.

\begin{figure}[htb]
\begin{center}
\begin{tabular}{cc}
\includegraphics[scale=0.33,clip=true,angle=0]{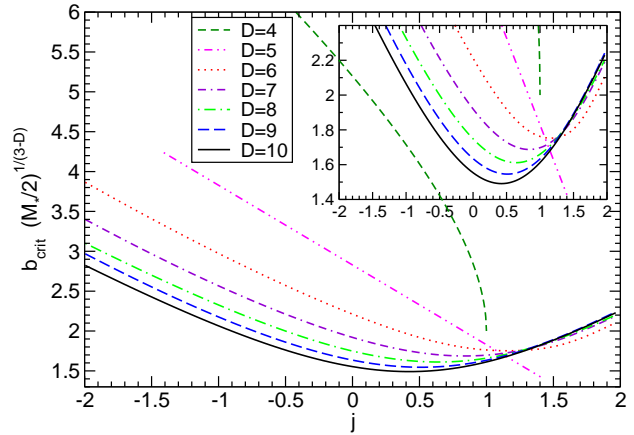}
\end{tabular}
\end{center}
\caption{\label{bc-j-MP} Critical impact parameter for $E = 10$ as a function
  of $j$ for Myers-Perry BHs in different dimensions $D$.  The inset zooms
  along the $y$-axis to show the local minima more clearly.}
\end{figure}

For dimensions higher than five, an asymptotic analysis is more involved due
to the higher inverse polynomial order of the effective potential. However, a
numerical calculation is straightforward.  In Fig.~\ref{bc-j-MP} we plot the
dimensionless impact parameter $b_{\rm crit}/(M_*/2)^{1/(D-3)}$ as a function
of $j$ for different values of $D$.  The calculations in Fig.~\ref{bc-j-MP}
refer to $E = 10$, but this is already a good approximation of the asymptotic
value of the impact parameter in the ultrarelativistic region. Observe that as
$D$ increases, the minimum in $b_{\rm crit}/(M_*/2)^{1/(D-3)}$ corresponds to
smaller values of $j$.

\begin{figure*}[htb]
\begin{center}
\begin{tabular}{cc}
\includegraphics[scale=0.5,clip=true]{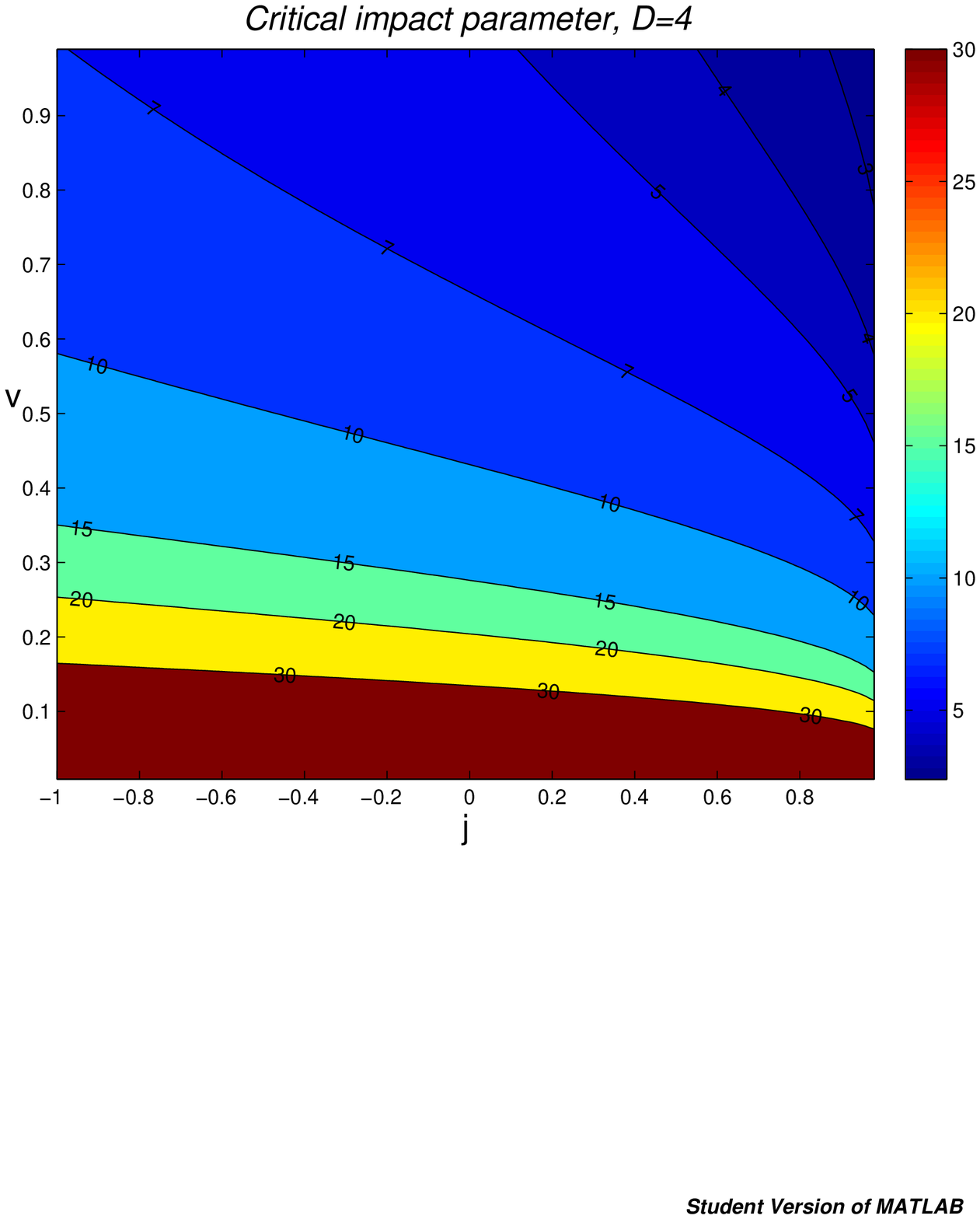} &
\includegraphics[scale=0.5,clip=true]{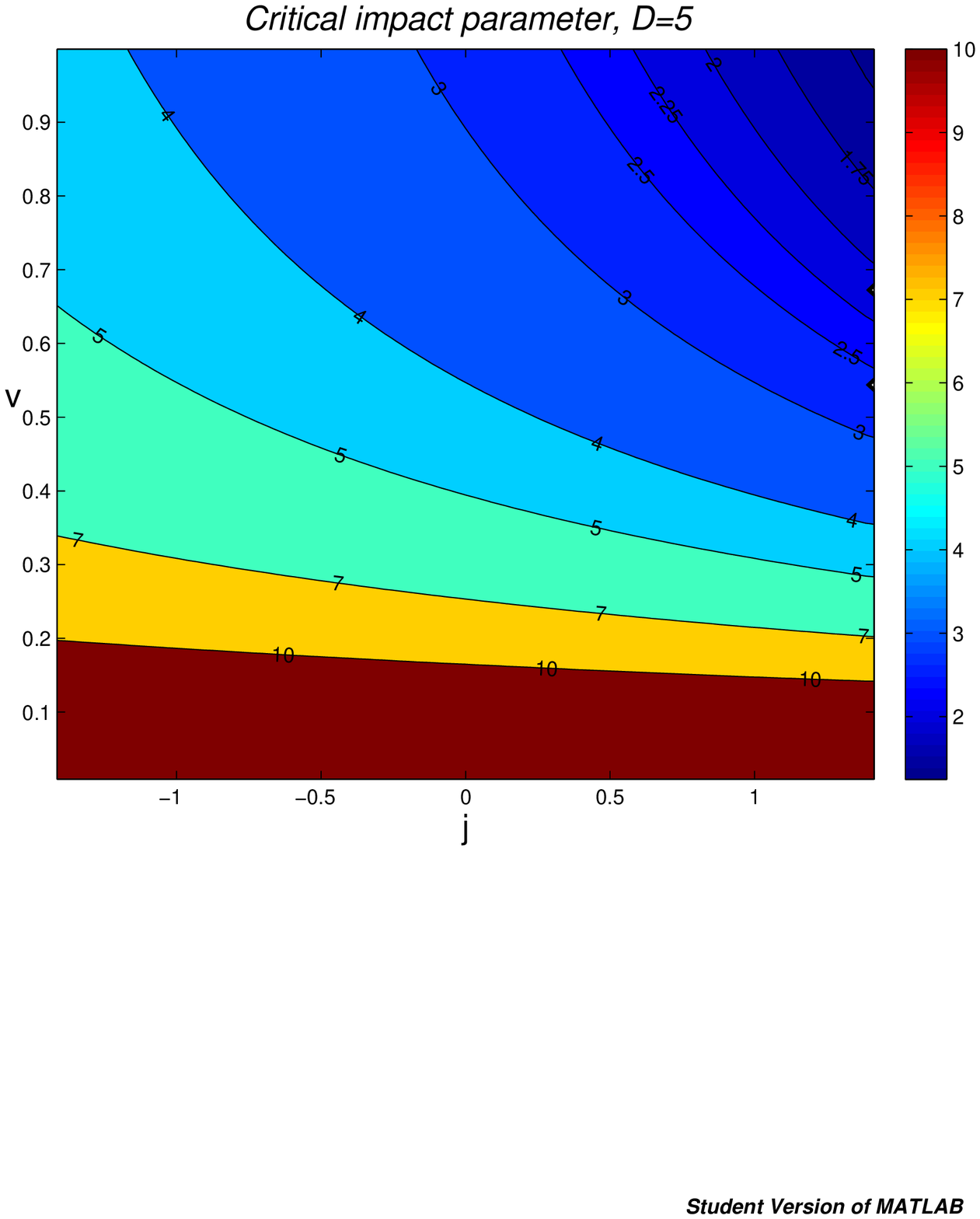} \\
\includegraphics[scale=0.5,clip=true]{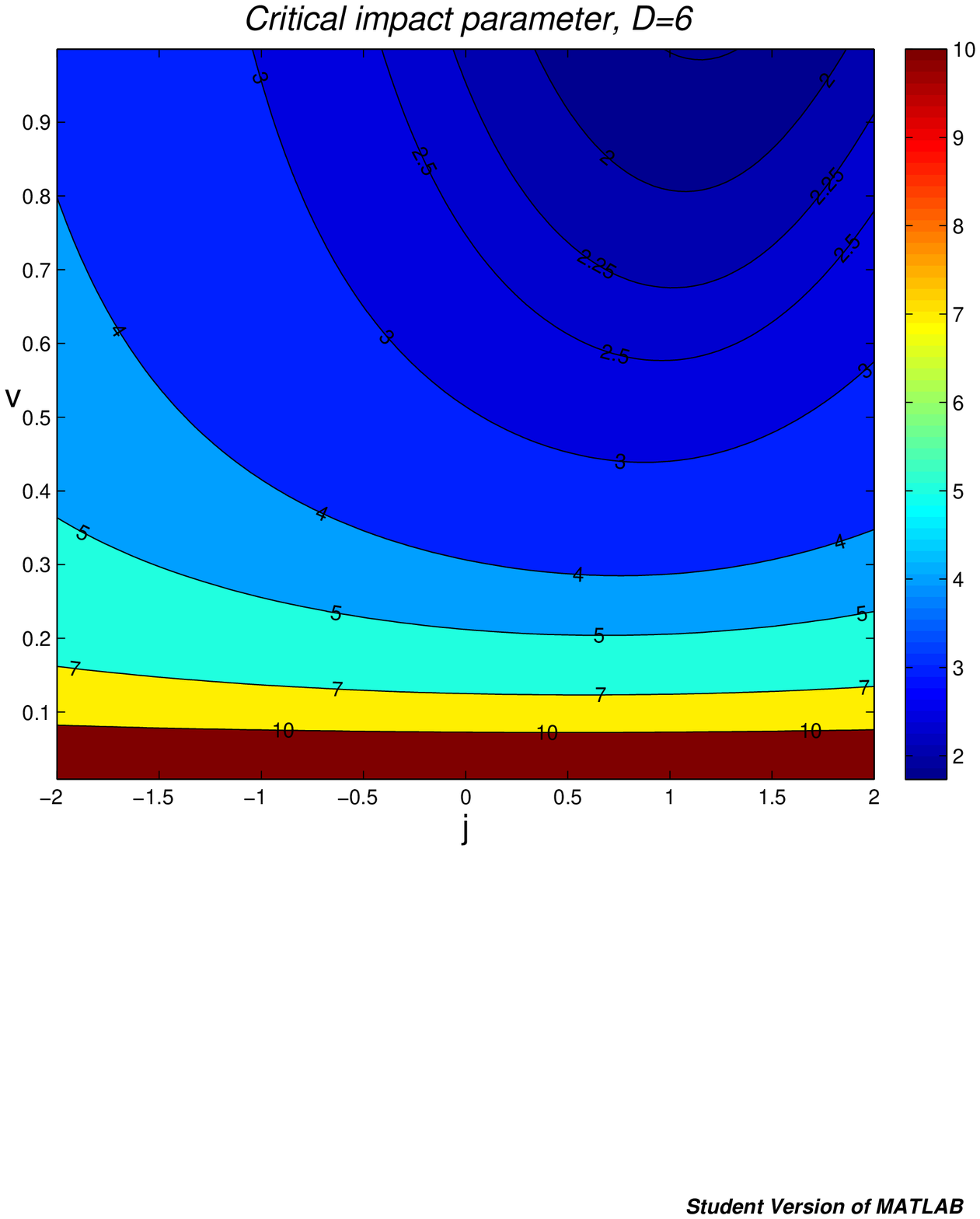} &
\includegraphics[scale=0.5,clip=true]{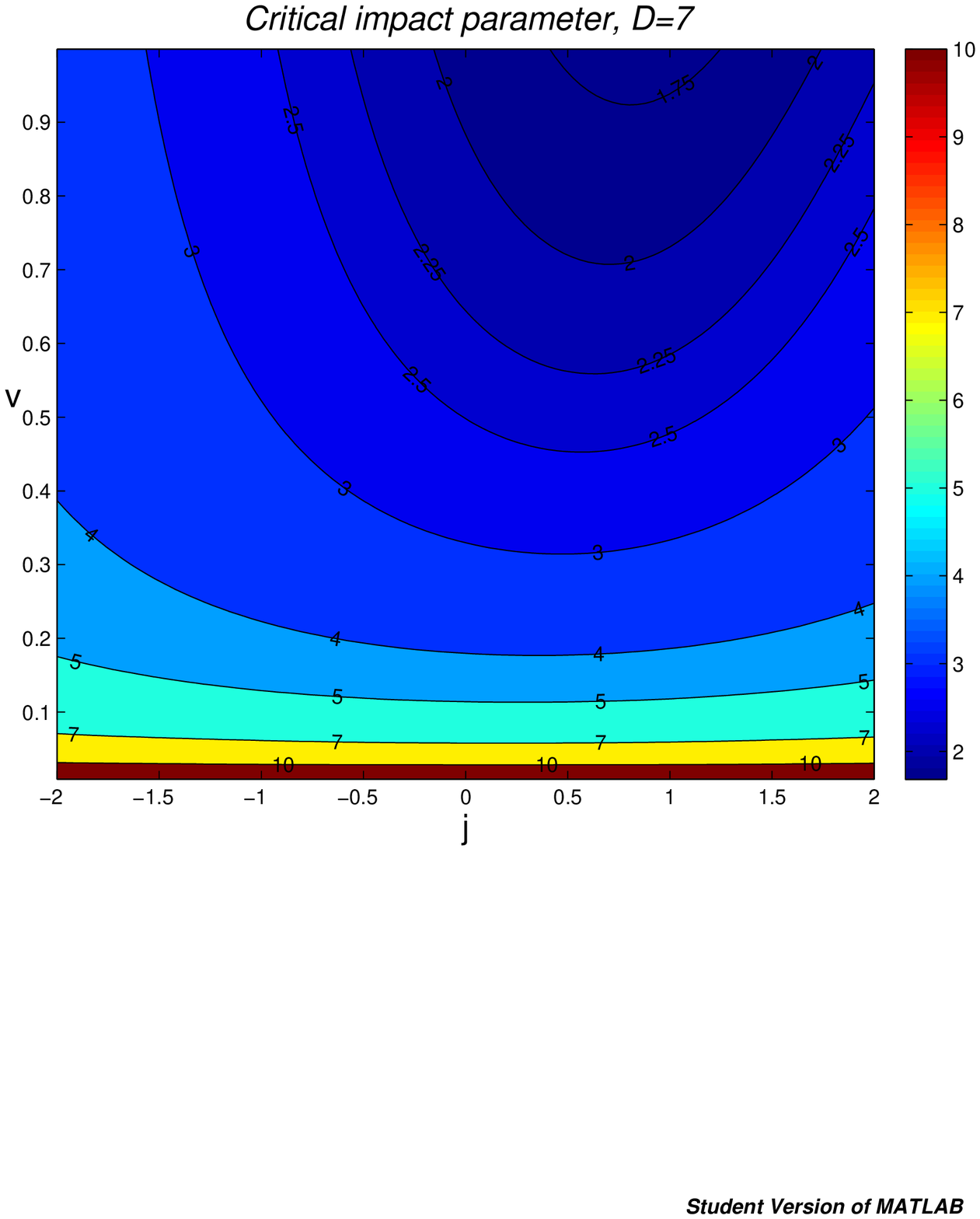} 
\end{tabular}
\end{center}
\caption{\label{bchighd} Contour plots of the critical impact parameter in the
  $(v\,,j)$ plane for corotating orbits ($j>0$) and counterrotating orbits
  ($j<0$) in dimensions $D=4$, $D=5$, $D=6$ and $D=7$.}
\end{figure*}

The $(v,j)$ phase space is explored in more detail in Fig.~\ref{bchighd},
where we show contour plots of the dimensionless impact parameter for both
corotating and counterrotating geodesics in different dimensions. Note that
the smallest critical impact parameter corresponds to corotating,
ultrarelativistic geodesics, but the BH must not necessarily be maximally
spinning when $D>5$. For $6\leq D\leq 10$ the critical impact parameter
presents a local minimum $b_{\rm crit}^{\rm min}$ as a function of $j$, which
is well described by the following quadratic fit:
\be 
b_{\rm crit}^{\rm min} \left(\frac{M_{*}}{2}\right)^{\frac{1}{3-D}} = 
3.978 - 0.679 D + 0.033 D^{2}\,.
\label{bcritmin}
\ee
This fit was constructed considering only spacetimes with $D\leq 10$, so
its extrapolation to $D>10$ should be used with caution.
%
%where the fitting parameters are $(c_{1},c_{2},c_{3}) = (3.978,-0.679,0.033)$.
%
% with $(c_{1},c_{2},c_{3}) = (3.978,-0.6795,0.03254)$.
% fit = 3.977726254261671 - 0.6795274022304891 d + 0.032541180789890316 d^2

%%%%%%%%%%%%%%%%%%%%%%%%%%%%%%%%%%%%%%%%%%%%%%%%%%%%%%%%%%%%%%%%%%%%%%%%%%%%%%%%%%%%%%%%%%%%%%%%%%%%%%%%%%%%%%%%%%%%%%%%%%%%%%%%%%%%%%%%%
\section{The ZFL in grazing collisions\label{zflgrazing}}
%%%%%%%%%%%%%%%%%%%%%%%%%%%%%%%%%%%%%%%%%%%%%%%%%%%%%%%%%%%%%%%%%%%%%%%%%%%%%%%%%%%%%%%%%%%%%%%%%%%%%%%%%%%%%%%%%%%%%%%%%%%%%%%%%%%%%%%%%

\begin{figure}
\begin{center}
\begin{tabular}{c}
\epsfig{file=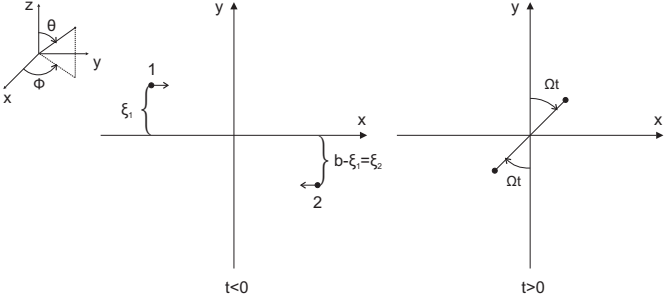,width=8.8cm,angle=0}
\end{tabular}
\end{center}
\caption{The system before and after the collision.  $\xi_1$ is defined in
  Eq.~(\ref{xi}).  }
\label{esquema}
\end{figure}

In this section we generalize the classic ZFL calculations for head-on
collisions \cite{Smarr:1977fy,Adler:1975dj} to the case of collisions with
finite impact parameter. The initial configuration consists of two point
particles with mass $M_k$ freely moving toward each other with constant,
positive velocity $v_k$, corresponding to boost factors $E_k=(1-v_k^2)^{-1/2}$
($k=1\,,2$). For convenience the axes are oriented such that the initial
motion is in the $x$--direction (see Fig.~\ref{esquema}). We assume that at
$x=0$ the particles ``collide'' with generic impact parameter $b$ and form a
single final body (strictly speaking this assumption is only valid for small
impact parameters, because we expect the bodies to scatter when $b$ is large
enough). Since the collision is not head-on (and since the energy loss is not
included in the motion of point particles), some confining force is necessary
to bind the particles. In fact, we show below that additional ``stresses'' are
required to guarantee energy conservation (cf. Ref.~\cite{Price:1973ns}).

Before the collision the particles have four-positions and four-momenta given
by:
\beq
x_{1}^{\mu}&=&(t,v_{1}t,\xi_1,0)\,,\qquad  \,\,\,
x_{2}^{\mu}=(t,-v_{2}t,-\xi_2,0)\,,\\
p_{1}^{\mu}&=&E_{1}M_{1} (1,v_{1},0,0)\,,\quad
p_{2}^{\mu}=E_{2}M_{2} (1,-v_{2},0,0)\,,\nn
\eeq
where $\xi_{1}$ ($-\xi_{2}$) is the projection of the position of particle 1
($2$) along the $y$--axis before the collision.  If the system's center of
mass is at $y = 0$, the position of particle 2 before the collision can also
be written as $y=-(b-\xi_1) \equiv -\xi_2$, where $b$ is the impact parameter.

At $t=0$ the particles become constrained to move as if they were attached to
an infinitesimally thin, massless rod of length $b$.  This fictional rod is an
idealization, but it is necessary to guarantee energy-momentum
conservation. For $t>0$ the particles remain attached to the rod, so that (in
the center-of-momentum frame) they rotate around the origin at fixed
separation $b$. Using primes to denote final states, the four-positions and
four-momenta after the collision are:
\beq
(x_{1}^{\mu})'&=&(t , \xi_1 S,\xi_1 C,0)\nn\,,\\
(x_{2}^{\mu})'&=&(t,-\xi_2 S,-\xi_2 C,0)\,,\nn\\
(p_{1}^{\mu})'&=&E_{1} M_{1}(1, \xi_1 \Omega C,-\xi_1 \Omega S,0)\,, \nn\\
(p_{2}^{\mu})'&=&E_{2} M_{2}(1,-\xi_2 \Omega C,\xi_2 \Omega S,0)\,,\nn
\eeq
where $S\equiv \sin{(\Omega t)}$, $C\equiv \cos{(\Omega t)}$.

A few relations can be derived between $(\xi_{1},\xi_{2})$ and $(v_{1},v_{2})$
due to the constraints we imposed. The requirement that the system's center of
mass is at $y=0$ implies
\beq
\xi_1=\frac{b E_{2} M_{2} }{E_{1} M_{1}
+E_{2} M_{2}}\,. \label{xi}
\eeq
In the center-of-momentum frame we also have
\be
E_1 M_1 v_1=E_2 M_2 v_2\,.
\ee

For an instantaneous collision the energy-momentum tensor of the system is
given by:
\beq 
\label{Tuv}
T^{\mu \nu}(\textbf{x},t)&=&\sum^2_{k=1}
\frac{p_k^{\mu}p_k^{\nu}}{E_k M_k}
\delta^{3}(\textbf{x}-\textbf{x}_k(t)) \Theta (-t) \\
&+& \sum^2_{k=1}
\frac{(p_k^{\mu})'(t)(p_k^{\nu})'(t)}{E_k M_k}
\delta^{3}(\textbf{x}-\textbf{x}'_k(t)) \Theta (t)\,, \nn
\eeq
where the boldface denotes a three-vector, and the angular momentum is
\beq
S_{3}&=&\int (x^{1}T^{20}-x^{2}T^{10}) d^{3}\textbf{x}\\
&=&-\sum_{k=1}^2
\left[
\Theta(-t) E_k M_k v_k \xi_k +
\Theta(t) E_k M_k \Omega \xi_k^2
\right]\,.
\nn
\eeq
Angular momentum conservation implies that the rotation frequency must be
\be
\Omega=\frac{E_{1} M_{1} v_{1} \xi_1 +
E_{2} M_{2} v_{2} \xi_2}{ E_{1} M_{1} \xi_1^2 + E_2 M_2 \xi_2^2}\,.
\label{rotfr}
\ee
If we naively take the stress-energy tensor of Eq.~\eqref{Tuv} to be the full
energy-momentum of the system, we would find that it is not covariantly
conserved, i.e.~$\nabla_{\mu} T^{\mu \nu} = 0$ for $\nu = t,z$ but
$\nabla_{\mu} T^{\mu \nu}\neq 0 $ for $\nu=x,y$.  In fact, one finds (in the
center-of-momentum frame)
\beq
\nabla_{\mu} T^{\mu x}&=&-E_1 M_1 \xi_1 \Omega^2 S \delta(x-\xi_1 S)
\delta(y-\xi_1 C) \delta(z) \Theta(t)\nn\\
&+&E_2 M_2 \xi_2 \Omega^2 S \delta(x+\xi_2 S)
\delta(y+\xi_2 C) \delta(z) \Theta(t) \, ,\nn\\
\nabla_{\mu} T^{\mu y}&=&-E_1 M_1 \xi_1 \Omega^2 C \delta(x-\xi_1 S)
\delta(y-\xi_1 C) \delta(z) \Theta(t)\nn\\
&+&E_2 M_2 \xi_2 \Omega^2 C \delta(x+\xi_2S)
\delta(y+\xi_2 C) \delta(z) \Theta(t) \,.\nn
\eeq
Physically, this nonconservation of stress-energy is due to neglecting the
energy-momentum associated with the fictitious rod that keeps the particles in
circular orbit. 

Energy-momentum conservation can be enforced by adding an additional term for
each particle that represents this constraining force. The contribution of
such forces to the gravitational radiation emitted by a particle in circular
orbit was studied by Price and Sandberg \cite{Price:1973ns}. By adding a
radial tension $\tau_k(r)$ for each particle and imposing that $\nabla_{\mu}
T^{\mu \nu}= 0$ we get the following contributions to the energy-momentum
tensor:
\beq
T^{xx}_{\rm tens}(t,\textbf{x})&=&
-S^2 \delta(\cos\theta) \Theta(t)
\sum_{k=1}^2 \tau_k(r) \delta(\phi+\Omega t-\phi_k)\,,
\nn\\
T^{yy}_{\rm tens}(t,\textbf{x})&=&
-C^2 \delta(\cos\theta) \Theta(t)
\sum_{k=1}^2 \tau_k(r) \delta(\phi+\Omega t-\phi_k)\,,
\nn\\
T^{xy}_{\rm tens}(t,\textbf{x})&=&
-SC \delta(\cos\theta) \Theta(t)
\sum_{k=1}^2 \tau_k(r) \delta(\phi+\Omega t-\phi_k)\,,
\nn
\eeq
where $\phi_1=\pi/2$, $\phi_2=3\pi/2$ and
\be
\tau_k(r)=\frac{M_k \xi_k \Omega^2 \Theta(\xi_k-r)}{r^2\sqrt{1-(\xi_k
    \Omega)^2}}\,,
\quad
(k=1,\,2)\,.
\ee
Here $r=\sqrt{x^2+y^2+z^2}$, $\theta$ is the polar angle measured from the
positive $z$-axis, and $\phi$ is the azimuthal angle in the $x$--$y$ plane
measured from the $x$-axis (see Fig.~\ref{esquema}). The factor
$\left[1-(\xi_k \Omega)^2\right]^{-1/2}$ is just the boost factor for particle
$k$ in circular motion with angular frequency $\Omega$. The stresses vanish
for $b=0$, as one would expect.

%%%%%%%%%%%%%%%%%%%%%%%%%%%%%%%%%%%%%%%%%%%%%%%%%%%%%%%%%%%%%%%%%%%%%%%%%%%%%%%%
\subsection{Equal-mass collisions\label{eqmassZFL}}
%%%%%%%%%%%%%%%%%%%%%%%%%%%%%%%%%%%%%%%%%%%%%%%%%%%%%%%%%%%%%%%%%%%%%%%%%%%%%%%%

In this subsection we study the equal-mass case $M/2 \equiv M_{1}=M_{2}$,
where $M$ is the total mass, and $v_1=v_2=v$, $E_1=E_2=E$.  According to
Eq.~(\ref{rotfr}), after the collision the particles are on a bound circular
orbit with radius $b/2$ and rotational frequency $\Omega=2v/b$.
The Fourier transform of the energy-momentum tensor (\ref{Tuv}) yields
\begin{widetext}
\beq
\label{ftransf}
T^{\mu \nu}(\textbf{k},\omega)&=&
\frac{p_{1}^{\mu}p_{1}^{\nu}}
{2\pi iE_{1}M_1(\omega - v_{1}k_{x})}
e^{-ik_{y} b/2}
+ \frac{p_{2}^{\mu}p_{2}^{\nu}}{2\pi i E_{2} M_2 (\omega + v_{2}k_{x})}
e^{ik_{y} b/2}+\\
&+&\sum_{k=1}^2 \int_{-\infty} ^{\infty}
\frac{p_k^{\prime\mu}(t)p_k^{\prime\nu}(t)}{2\pi E_k M_k}
\exp(i \omega t-i \textbf{k}\cdot\textbf{x}'_k(t) ) \Theta(t) dt+
\frac{1}{2\pi}\int d^4x\, T^{\mu\nu}_{\rm tens}(\textbf{x},t)
e^{i \omega t-i \textbf{k}\cdot\textbf{x}}\,,\nn
\eeq
\end{widetext}
where $d^4x=dx\,dy\,dz\,dt$ and $\mathbf{k}$ is the wave vector:
\be
k_{x}=\omega \sin\phi \cos\theta\,,\,\,
k_{y}=\omega \sin\phi \sin\theta \,,\,\,
k_{z}=\omega \cos\phi\,.
\ee
We also have
\be
e^{i \omega t-i \mathbf{k}\cdot\mathbf{x}'_k(t) }
=e^{i \omega t}
\exp\left(i 
\lambda_k \frac{\omega b}{2}\sin\theta \sin(\Omega t+\phi)
\right)\,,
\ee
where $k=1,\,2$ is the particle index, $\lambda_1=-1$ and $\lambda_2=1$. If we
set $\alpha=\Omega t+\phi$ and $2\eta_k=\lambda_k \omega b\sin\theta$
the last exponential can be written in terms of Bessel functions of the first
kind, using the Jacobi-Anger expansion \cite{Abramowitz:1970as}:
\be
e^{i \eta \sin \alpha }=\sum_{n=-\infty}^{n=+\infty} J_n (\eta) e^{i n \alpha}\,.
\ee
For large $n$ the Bessel functions satisfy \cite{Abramowitz:1970as}
\be
J_n(\eta)\sim \frac{1}{\sqrt{2\pi n}}\left(\frac{e\eta}{2n}\right)^n\,.
\ee
A time-integration introduces an additional factor of $1/n$, so the series
converges rapidly for large $|n|$ and we can truncate it at some moderately
large value of $n=N$ to get an accurate approximation of the
integral\footnote{Actually, the series should be approximated by summing from
  $n=n_0-N$ to $n=n_0+N$, where $n_0$ is the value of $n$ which maximizes the
  absolute value of the terms being summed. After the integration terms of the
  form $1/(\omega - n \Omega)$ appear. This means that the largest
  contribution to the sum corresponds to some $n_0\neq 0$.  However it can be
  checked that $N\gg n_0$ for the range of parameters considered here, so the
  sum can be taken in a symmetric interval around 0.}.
Typically, $N\gtrsim 10$ is sufficient for an accuracy of $1\%$ or better.

The integration of the stresses proceeds in a similar way. After integrating
in $\theta$ and $\phi$, the same Bessel function expansion can be used for the
time-integration. The integral of Bessel functions with respect to $r$ can be
evaluated using the following identity \cite{Abramowitz:1970as}:
\be
\int_0^\eta J_\nu (r)dr=2\sum_{k=0}^\infty J_{\nu+2k+1}(\eta),\;\quad 
{\rm Re}(\nu)>-1\,.
\ee

%%%%%%%%%%%%%%%%%%%%%%%%%%%%%%%%%%%%%%%%%%%%%%%%%%%%%%%%%%%%%%%%%%%%%%%%%%%%%%%
\subsubsection{Radiation Spectrum}
%%%%%%%%%%%%%%%%%%%%%%%%%%%%%%%%%%%%%%%%%%%%%%%%%%%%%%%%%%%%%%%%%%%%%%%%%%%%%%%

The energy per solid angle and per unit frequency emitted in the direction
$\mathbf{\hat{k}}=\mathbf{k}/\omega$ is \cite{Weinberg}:
\be
\frac{d^{2}E}{d\omega d\Omega}=
2\omega^{2}\left(T^{\mu\nu}(\mathbf{k},\omega)T^{*}_{\mu\nu}(\mathbf{k},\omega)
-\frac{1}{2}\left|T^{\lambda}_{\;\;\;\lambda}(\mathbf{k},\omega)\right|^{2}\right)
\,, \label{energy}
\ee
where the asterisk stands for complex conjugation.  The energy can also be
expressed in terms of the purely spacelike components of $T^{\mu\nu}$. The
conservation equation for $T^{\mu\nu}$ implies that $k_{\mu} T^{\mu
  \nu}(\mathbf{k},\omega)=0$, so it is possible to write $T^{00}$ and $T^{0i}$
in terms of $T^{ij}$:
\beq
T_{00}(\mathbf{k},\omega)&=&\hat{k}^i \hat{k}^j T_{ij}(\mathbf{k},\omega)\,,\\
T_{0i}(\mathbf{k},\omega)&=&-\hat{k}^j T_{ij}(\mathbf{k},\omega)\,.
\eeq
With these identities at hand, Eq.~(\ref{energy}) can be written as:
\be
\frac{d^2 E}{d\omega d\Omega}=
2 \omega^2 \Lambda_{ijlm}(\hat{k})
T^{*ij}(\mathbf{k},\omega) T^{lm}(\mathbf{k},\omega) \,,
\label{energyspace}
\ee
where following Ref.~\cite{Weinberg} we defined
\beq
\Lambda_{ijlm}(\hat{k})&=&\delta_{il}\delta_{jm}
-2\hat{k}_j \hat{k}_m \delta_{il}
+\frac{1}{2}\hat{k}_i \hat{k}_j \hat{k}_l \hat{k}_m
-\frac{1}{2}\delta_{ij}\delta_{lm} \nn\\
&+&\frac{1}{2}\delta_{ij} \hat{k}_l \hat{k}_m
+ \frac{1}{2} \delta_{lm}\hat{k}_i \hat{k}_j\,.
\eeq
\begin{figure}
\begin{center}
\includegraphics[scale=0.30,clip=true]{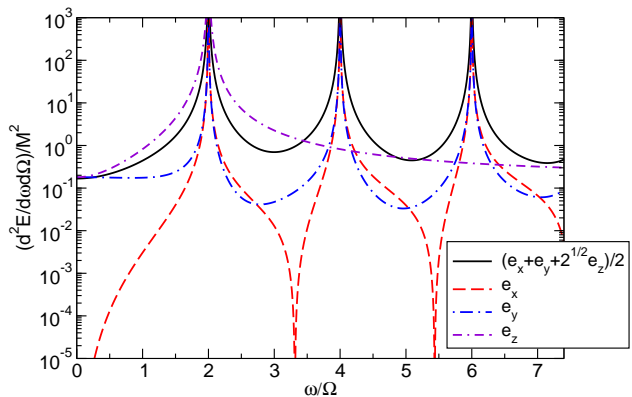} 
\end{center}
\caption{Energy per unit solid angle and per unit frequency emitted in the
  directions $\mathbf{\hat{k}}=\mathbf{e_x}\,,\mathbf{e_y}, \mathbf{e_z},
  (\mathbf{e_x}+\mathbf{e_y}+\sqrt{2}\mathbf{e_z})/2$ by equal-mass binaries
  with $E=3$, as a function of $\omega/\Omega$.}
\label{fig:spectraequalmass}
\end{figure}

In Fig.~\ref{fig:spectraequalmass} we plot the energy spectrum for $E=3$ along
four different directions:
$\mathbf{\hat{k}}=\mathbf{e_x}\,,\mathbf{e_y}\,,\mathbf{e_z}\,,
(\mathbf{e_x}+\mathbf{e_y}+\sqrt{2}\mathbf{e_z})/2$. For dimensional reasons,
there is no need to fix $b$ as long as the energy is plotted as a function of
$\omega/\Omega$. All spectra diverge when $\omega=2\Omega$, as expected of a
rigid symmetric body rotating with angular frequency $\Omega$. For
$\mathbf{\hat{k}}=\mathbf{e_z}$ the spectrum only diverges at $\omega=2\Omega$
(see Ref.~\cite{Poisson:1993vp} for a discussion of particles in circular
orbit in the Schwarzschild geometry), but in all other directions the spectrum
diverges at even multiples of the rotational frequency $\Omega$. The same
qualitative features hold for higher boost parameters.

%%%%%%%%%%%%%%%%%%%%%%%%%%%%%%%%%%%%%%%%%%%%%%%%%%%%%%%%%%%%%%%%%%%%%%%%%%%%%%%%
\subsubsection{Head-on collisions}
%%%%%%%%%%%%%%%%%%%%%%%%%%%%%%%%%%%%%%%%%%%%%%%%%%%%%%%%%%%%%%%%%%%%%%%%%%%%%%%%

Let us consider the $b=0$ limit, which corresponds to a head-on collision and
for which we can compare against known results \cite{Smarr:1977fy,Adler:1975dj}. 
In this limit, the only nonvanishing components of the energy-momentum tensor are:
\beq
2\pi \omega T^{tt}(\mathbf{k},\omega)&=& i E M
-\frac{i E M}{1-v^2 \sin^2 \theta  \cos^2\phi}\,,\nn\\
2\pi \omega T^{tx}(\mathbf{k},\omega)&=&
-\frac{i E M v^2 \sin \theta \cos \phi }
{1 - v^2\sin^2\theta\cos^2\phi}\,,  \nn\\
2\pi \omega T^{xx}(\mathbf{k},\omega)&=&
-\frac{i E M v^2 }
{1 - v^2\sin^2\theta\cos^2\phi}\,.\nn
\eeq
The energy spectrum per unit solid angle is then given by
\be
\frac{d^2 E}{d\omega d\Omega}=
\frac{ E^2 M^2 v^4 \left(\sin ^2\theta \cos ^2\phi-1\right)^2}
{4 \pi^2 \left(v^2 \sin ^2\theta  \cos^2\phi-1\right)^2} \,.
\label{b0}
\ee
This agrees with previous results in the literature. Indeed, after a trivial
redefinition of angles, Eq.~(\ref{b0}) is equal to Eqs.~(2.19) and (2.12) of
Refs.~\cite{Smarr:1977fy} and \cite{Adler:1975dj}, respectively. In
Refs.~\cite{Smarr:1977fy,Adler:1975dj} the angular variable $\theta$ is the
angle between the radiation direction and the momenta of the particles; the
substitution of their $\cos\theta$ by $\sin\theta\cos\phi$ yields
Eq.~(\ref{b0}).  For ease of comparison with NR results
\cite{Sperhake:2008ga,Sperhake:2009jz}, it is convenient to expand the above
expression in spin-weighted spherical harmonics with spin weight $s=-2$. Such
an expansion is discussed in Appendix~\ref{app:multipoles}.

Recalling that $\mathbf{e_x}$ corresponds to $(\theta=\pi/2\,,\phi=0)$,
$\mathbf{e_y}$ corresponds to $(\theta=\pi/2\,,\phi=\pi/2)$ and $\mathbf{e_z}$
corresponds to $\theta=0$ we get:
\begin{subequations}
\beq
\frac{d^2 E}{d\omega d\Omega}&=&0 
%\hskip18mm 
\quad
{\rm along} \,\,\mathbf{e_x}\,,\\
\frac{d^2 E}{d\omega d\Omega}&=&\frac{ E^2 M^2 v^4}
{4 \pi^2} 
\quad 
{\rm along} \,\,\mathbf{e_y},\,\mathbf{e_z}\,.\label{b01}
\eeq
\label{b0eqns}
\end{subequations}

The radiated momentum per unit frequency for this head-on collision is found
by the following integral over a two-sphere at infinity, $S_\infty$, centered
on the coordinate origin:
\be 
\frac{dP_i}{d\omega}=
\int_{S_\infty} \frac{d^2E}{d\omega d\Omega} n_i d\Omega\, , \label{radmom}
\ee
where $n_i$ is a unit radial vector normal to $S_{\infty}$. In the present
case we find that the radiated momentum vanishes, i.e. ${dP_i}/{d\omega} = 0$,
as one would expect.

%%%%%%%%%%%%%%%%%%%%%%%%%%%%%%%%%%%%%%%%%%%%%%%%%%%%%%%%%%%%%%%%%%%%%%%%%%%%%%%%
\subsubsection{Zero-frequency limit}
%%%%%%%%%%%%%%%%%%%%%%%%%%%%%%%%%%%%%%%%%%%%%%%%%%%%%%%%%%%%%%%%%%%%%%%%%%%%%%%%

For arbitrary impact parameters our results show that, in the limit
$b\omega\to 0$, the energy spectrum is independent of $b$ and given by
Eq.~(\ref{b0}). This is of course consistent with the head-on results of Smarr
\cite{Smarr:1977fy} and Adler and Zeks \cite{Adler:1975dj}. Numerical
calculations support this conclusion and reveal additional details for small
but nonzero frequencies. The stress terms give the following contributions to
the energy-momentum tensor:
\beq
\omega T^{xx}_{\rm tens}(\mathbf{k},\omega) \Big| _{\omega =0}=\omega
T^{yy}_{\rm tens}(\mathbf{k},\omega)\Big| _{\omega =0}&=&
-\frac{i b^2 E M \Omega ^2}{16 \pi}\,,
\nn\\
\omega T^{xy}_{\rm tens}(\mathbf{k},\omega)\Big| _{\omega =0}&=&0 \label{ten0}\,.
\eeq
For $\omega=0$ the constraining forces provide a nonvanishing contribution to
the energy-momentum tensor. It is this particular contribution that allows one
to recover the ZFL of the energy spectrum, Eq.~(\ref{b0}), for {\it any}
impact parameter. This is one of the most intriguing results of this incursion
into the properties of the ZFL for collisions with nonzero impact parameter.

\begin{figure}[htb]
\begin{center}
\includegraphics[scale=0.33,clip=true]{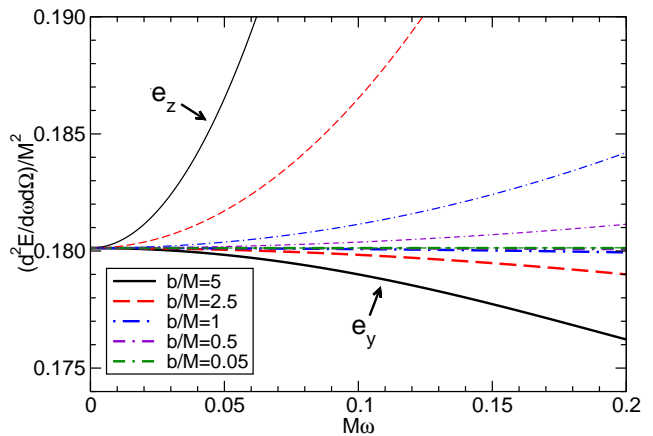}
\end{center}
\caption{Normalized energy spectrum per unit solid angle emitted in the
  directions $\mathbf{\hat{k}}=\mathbf{e_y}$ (thick lines) and
  $\mathbf{\hat{k}}=\mathbf{e_z}$ (thin lines) as a function of $M \omega$ for
  several values of $b/M$ (as indicated in the legend) and $E=3$. As indicated
  by Eq.~\eqref{b0eqns} the ZFL for these two different directions is the same
  and approximately equal to $0.18013 M^2$. \label{by}}
\end{figure}

The spectra for small frequencies along the directions
$\mathbf{\hat{k}}=\mathbf{e_y}$ and $\mathbf{\hat{k}}=\mathbf{e_z}$ are
plotted in Fig.~\ref{by} for different values of $b/M$. All spectra have the
same ZFL, as discussed above.  Specializing Eq.~\eqref{b01} to the case $E=3$
we get $\frac{d^2 E}{d\omega d\Omega}\Big| _{\omega =0}=0.18013 M^2$ for
$\mathbf{e_y},\mathbf{e_z}$.  This is in very good agreement with the
numerical results shown in Fig.~\ref{by}.

For small but finite frequencies, we find that the slope of the energy
spectrum depends on direction. It is positive (negative) for
$\mathbf{\hat{k}}=\mathbf{e_y}$ ($\mathbf{e_z}$, respectively), and it
increases with $b/M$.  An expansion of the energy for small $b\omega$ yields
\begin{widetext}
\beq
\frac{d^2E}{d\omega d\Omega}&=&
\frac{ E^2 M^2 v^2}
{16\pi^2 \left(1-v^2 \sin ^2\theta \cos ^2\phi\right)^2}
\bigg\{ 4 v^2 (\sin^2 \theta \cos^2 \phi -1)^2- \nn \\
 &-&\frac{1}{6}\Big[
v^2 \left(8 v^2+3\right)
\sin ^6\theta \cos ^4\phi+\sin ^2\theta
\left(\left(8 v^2+3\right) \cos2 \phi+12 \left(v^2+1\right)\right)- \nn \\
&-&\sin ^4\theta \cos ^2\phi
\left(6 v^4+\left(2 v^2+3\right) v^2 \cos 2 \phi+20 v^2+3\right)
-12\Big]
\left(b\, \omega\right)^2 \bigg\}+ \mathcal{O}\left[(b\,\omega)^3 \right]\,.
\label{aproxwpeq}
\eeq
\end{widetext}

Thus, within our model the spectrum typically has quadratic corrections,
except along $\hat{\mathbf{k}}=\mathbf{e_x}$, in which case the first
nonvanishing contribution to the energy spectrum is of order
$(b\,\omega)^4$. In fact, we find
\be
\frac{d^2E}{d\omega d\Omega}
\bigg| _{\mathbf{\hat{k}}=\mathbf{e_x}}=-
\frac{ E^2 M^2 \left(3-2 v^2\right)^2}
{9216 \pi^2 }(b\,\omega)^4+
\mathcal{O}\left[(b\,\omega)^5 \right]\,.
\ee
These analytical expressions are in good agreement with the results shown in
Fig.~\ref{by}.

%%%%%%%%%%%%%%%%%%%%%%%%%%%%%%%%%%%%%%%%%%%%%%%%%%%%%%%%%%%%%%%%%%%%%%%%%%%%%%%%
\subsection{Extreme-mass ratio collisions}
%%%%%%%%%%%%%%%%%%%%%%%%%%%%%%%%%%%%%%%%%%%%%%%%%%%%%%%%%%%%%%%%%%%%%%%%%%%%%%%%

We now study collisions for $\mu \equiv M_1\ll M_2 \equiv M$ (the qualitative
features of the radiation for {\it generic} mass ratio are very similar to the
extreme-mass ratio case).  The energy spectrum can be computed in the
center-of-momentum frame. Since particle $2$ is much heavier than particle
$1$, particle $2$ is practically at rest in this frame, although we shall not
neglect its motion when we compute the energy-momentum tensor. Therefore, we
let $v_1\equiv v \gg v_2$ and $E_1\equiv E$.

From Eqs.~(\ref{xi}) and (\ref{rotfr}), the angular frequency and position of
particle $1$ are given by
\be
\Omega =\frac{E_1 \mu v_1 \xi_1 + E_2 M v_2 \xi_2}
{E_1 \mu \xi_1 + E_2 M \xi_2}\,,\quad
\xi_1 = \frac{b E_2 M}{E_1 \mu+ E_2 M}\,.
\ee
Once again we must add the stresses needed to constrain the particles in their
orbits, in order to have a conserved energy-momentum tensor.

%%%%%%%%%%%%%%%%%%%%%%%%%%%%%%%%%%%%%%%%%%%%%%%%%%%%%%%%%%%%%%%%%%%%%%%%%%%%%%%%
\subsubsection{Radiation Spectrum}
%%%%%%%%%%%%%%%%%%%%%%%%%%%%%%%%%%%%%%%%%%%%%%%%%%%%%%%%%%%%%%%%%%%%%%%%%%%%%%%%

\begin{figure}[ht]
\includegraphics[scale=0.33,clip=true]{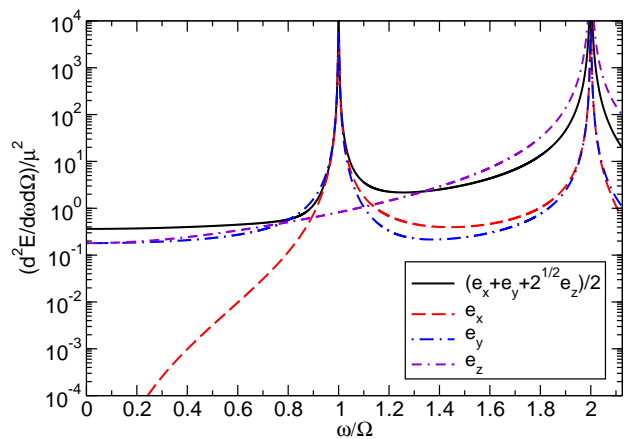}
\caption{Normalized energy spectrum per solid angle emitted in the directions
  $\mathbf{\hat{k}}=\mathbf{e_x},\mathbf{e_y},\mathbf{e_z},
  (\mathbf{e_x}+\mathbf{e_y}+\sqrt{2}\mathbf{e_z})/2$ as a function of
  $\omega/\Omega$ in the extreme-mass ratio case.
%Here $\Omega=v/b=0.471405/M$, to leading order in $\mu/M$.
  \label{mMx}}
\end{figure}

We compute the radiated energy using Eq.~(\ref{energy}) or
Eq.~(\ref{energyspace}). We expand the energy-momentum tensor in powers of
$\mu/M$, and compute the energy keeping only leading-order contributions in
$\mu/M$. A calculation of the radiation for $E=3$ (in the center-of-momentum
frame) along several different directions yields the spectra shown in
Fig.~\ref{mMx}.

The extreme-mass ratio configuration loses the angular symmetry of the
equal-mass case.  Therefore, the spectra now diverge for all multiples of
$\Omega$ for $\mathbf{\hat{k}}=\mathbf{e_x}$, $\mathbf{\hat{k}}=\mathbf{e_y}$.
The behavior is similar for other directions. For
$\mathbf{\hat{k}}=\mathbf{e_z}$, the spectrum only diverges for
$\omega=2\Omega$, in agreement with Poisson's findings for particles in
circular orbit around BHs \cite{Poisson:1993vp}.

%%%%%%%%%%%%%%%%%%%%%%%%%%%%%%%%%%%%%%%%%%%%%%%%%%%%%%%%%%%%%%%%%%%%%%%%%%%%%%%%
\subsubsection{Head-on collisions}
%%%%%%%%%%%%%%%%%%%%%%%%%%%%%%%%%%%%%%%%%%%%%%%%%%%%%%%%%%%%%%%%%%%%%%%%%%%%%%%%

For extreme-mass ratio head-on collisions ($b=0$) we find
\be
\frac{d^{2}E}{d\omega d\Omega}=
\frac{ E^2 \mu^2 v^4 \left(\sin^2\theta \cos^2\phi -1\right)^2}
{4 \pi^2 (v \sin\theta \cos \phi-1)^2}\,.
\label{mMb}
\ee
This expression coincides, as it should, with Eq.~(2.17) of
Smarr \cite{Smarr:1977fy}, once we take into account the different convention
on angles by an appropriate redefinition of angular variables (cf.~the
discussion following Eq.~\eqref{b0}.) For ease of comparison with perturbative
results of point particles in BH spacetimes, in Appendix \ref{app:multipoles}
we compute analytically the multipolar decomposition of this ZFL result in
spin-weighted spherical harmonics.

By computing the radiated momentum using Eq.~(\ref{radmom}) we find that it
vanishes along the $y$-- and $z$--axes, and that along the $x$--axis it is
given by
\be
\frac{dP^x}{d\omega}=
\frac{\mu^2 E^2 \left[v \left(15-13 v^2\right)-3 
\left(v^4-6 v^2+5\right) \text{arctanh}v\right]}{3 \pi  v^2 }\,.
\label{zflmomentum}
\ee
This result is in good agreement with the linear momentum radiated by point
particles falling into Schwarzschild BHs, as we will see in
section~\ref{sec:lm}.
%

%%%%%%%%%%%%%%%%%%%%%%%%%%%%%%%%%%%%%%%%%%%%%%%%%%%%%%%%%%%%%%%%%%%%%%%%%%%%%%%%
\subsubsection{Zero-frequency limit}
%%%%%%%%%%%%%%%%%%%%%%%%%%%%%%%%%%%%%%%%%%%%%%%%%%%%%%%%%%%%%%%%%%%%%%%%%%%%%%%%

Let us now consider the ZFL for generic values of the impact parameter.  As
$b\,\omega\to 0$ we find once again that the energy spectrum is independent of
the impact parameter (as it was for the equal-mass collisions of
section~\ref{eqmassZFL}).  The leading-order expression of the energy in
powers of $\mu/M$ is given by Eq.~\eqref{mMb}, reproducing Smarr's result for
head-on collisions. Including higher powers of $b\,\omega$ we get
\begin{widetext}
\beq
\frac{d^2E}{d\omega d\Omega}&=&
\frac{ E^2 \mu^2 v^2}{4\pi^2 
(v \sin\theta \cos\phi-1)^2}
\bigg\{
v^2 (\sin^2\theta \cos^2 \phi -1)^2 -
\\ \nn
&-&\frac{1}{192}
\Big[ (-4 v \left(2 v^2+3\right) \sin ^3\theta (\cos 2 \theta+3)
\cos 3 \phi+8 \sin^2\theta \cos2 \phi \left(\left(3-10 v^2\right)
\cos2 \theta-6 v^2+9\right)+\\ \nn
&+& v \sin\theta \cos\phi \left(\left(372-8 v^2\right)
\cos 2 \theta+\left(6 v^2+9\right) \cos 4 \theta+2 v^2+387\right)
+8 \left(2 v^2-21\right) \cos 2 \theta +\\ \nn
&+&\left(20 v^2-6\right) \cos4 \theta-6 \left(6 v^2+35\right)\Big]
(b\,\omega)^2 \bigg\}+\mathcal{O}\left[(b\,\omega)^3\right]\,.
\eeq
\end{widetext}
As in the equal-mass case, here too the radiation is suppressed along the
$x$--axis, where the leading contribution is of order $(b\,\omega)^4$:
\be
\frac{d^2E}{d\omega d\Omega} \bigg| _{\hat{k}=\mathbf{e_x}}=
\frac{ E^2 \mu^2 (3-2 v (v+3))^2}
{576 \pi^2}(b\,\omega) ^4
+\mathcal{O}\left[(b\,\omega)^5\right]\,.
\ee
%

%%%%%%%%%%%%%%%%%%%%%%%%%%%%%%%%%%%%%%%%%%%%%%%%%%%%%%%%%%%%%%%%%%%%%%%%%%%%%%%%
\subsection{Generality of the model}
%%%%%%%%%%%%%%%%%%%%%%%%%%%%%%%%%%%%%%%%%%%%%%%%%%%%%%%%%%%%%%%%%%%%%%%%%%%%%%%%

The most important result of our ZFL calculation for collisions with generic
impact parameter is perhaps that the ZFL itself is {\it independent} of the
impact parameter.  One of the limitations of the present calculation is that
the modeling of the collision is rather ad-hoc, especially in the
specification of the nature of the constraining forces. It is natural to ask
how the results would change if the constraining forces were modeled
differently. For example, in our toy model the final system consists of two
particles bound in a circular orbit, so the radiation spectrum shows peaks
typical of the radiation produced by rotating bodies. There is a chance that
the divergence at harmonics of the rotational frequency of the final system
could contaminate the low-frequency behavior of the spectrum.
%Thus, the dependence of the ZFL on the chosen description of the system must
%be studied.

To investigate this possibility, instead of considering the collision of two
point particles, we studied a point particle colliding with a special extended
matter distribution: specifically, we considered an infinitely thin, slowly
rotating, uniform disk. For brevity we do not report details of this
calculation here. Our main finding is that, if the disk is initially slowly
rotating (so that after the collision the system is at rest), the ZFL is the
{\it same} as in the case of two colliding particles. This is by no means a
proof that the ZFL is completely independent of the way one models the system.
It is however a hint that (as physical intuition would suggest) the ZFL should
depend only on the asymptotic momenta of the colliding particles.

%%%%%%%%%%%%%%%%%%%%%%%%%%%%%%%%%%%%%%%%%%%%%%%%%%%%%%%%%%%%%%%%%%%%%%%%%%%%%%%%
\section{Ultrarelativistic infall of point particles\label{ppart}}
%%%%%%%%%%%%%%%%%%%%%%%%%%%%%%%%%%%%%%%%%%%%%%%%%%%%%%%%%%%%%%%%%%%%%%%%%%%%%%%%

The formalism discussed in section~\ref{zflgrazing} is a flat-space
approximation valid for the low-frequency part of the energy spectrum and of
the gravitational waveforms. In this section we compute the radiation from the
linearized field equations in the {\it curved} background of a Schwarzschild
BH. This is an accurate description at all frequencies in the limit where one
of the binary components is much more massive than the other.

There is extensive literature on the gravitational radiation emitted by
particles following geodesics in BH backgrounds (see \cite{Nakamura:1987zz}
and Appendix C of \cite{Berti:2007fi} for summaries). Here we present an
incomplete overview of this literature. Davis {\it et al.} first studied the
radiation emitted by particles falling radially from rest into a Schwarzschild
BH \cite{Davis:1971gg} and the synchrotron radiation emitted by particles in
circular orbits \cite{Davis:1972dm} (see also \cite{Detweiler:1978ge} and
Detweiler's contribution to Ref.~\cite{Smarr}). Oohara and Nakamura computed
the energy, angular momentum and linear momentum radiated by particles falling
from rest with generic angular momentum into Schwarzschild BHs
\cite{Oohara:1984bw,Oohara:1983gq}. This work was later generalized to
particles on {\em scattering} orbits in Schwarzschild, starting either from
rest \cite{Oohara:1984ck} or with finite energy at infinity
\cite{Oohara:1984iu}.
Radial infalls into a nonrotating BH with finite energy were considered in
Refs.~\cite{Ruffini:1973ky,Ferrari:1981dh,Lousto:1996sx}, and the
ultrarelativistic limit was compared with Smarr's ZFL in
Refs.~\cite{Cardoso:2002ay,Cardoso:2002yj,Cardoso:2002jr} (in four dimensions)
and \cite{Berti:2003si} (in dimensions $D\geq 4$).  Oohara, Kojima and
Nakamura studied orbits plunging
\cite{Oohara:1984bw,Kojima:1983ua,Kojima:1984pz,Kojima:1984cj} and scattering
\cite{Kojima:1984cc} from rest in the case of rotating (Kerr) BHs. More recent
studies focused on the threshold of immediate merger using the geodesic
analogy \cite{Merrick:2007,Pretorius:2007jn,Grossman:2008yk,Healy:2009zm}.

Perhaps because of the limited astrophysical relevance of infalls with finite
energy at infinity, to our knowledge there is no detailed study of the
radiation emitted by point particles falling with generic energy and impact
parameter into Schwarzschild BHs. One purpose of this section is to fill this
surprising gap in the literature. A complementary study of scattering orbits
with generic energy can be found in Ref.~\cite{Oohara:1984iu}. The
generalization of this study to particles falling with arbitrary energy and
impact parameter into Kerr BHs is in preparation.

\begin{figure}
\includegraphics[scale=0.33,clip=true]{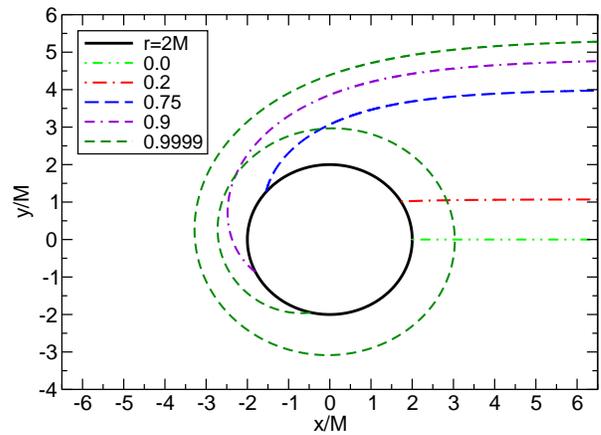}
\caption{\label{traj} Trajectories for different values of $L_z/L_{\rm crit}$
  (as indicated in the legend) and $E=3$. The black circle of radius $2$ marks
  the location of the horizon.}
\end{figure}

The radiation can be determined from the knowledge of the Sasaki-Nakamura
wave function $X_{lm}$, which (in the frequency domain) can be written in the
form
\be 
\frac{d^2X_{lm}}{dr_*^2} +
\left[
\omega^2-\frac{\Delta}{r^5}\left(l(l+1)r-6M\right)
\right]X_{lm}=S_{lm}\,.
\label{SNeq}
\ee
Here $(l,\,m)$ are (tensor) spherical harmonic indices resulting from a
separation of the angular variables, $\omega$ is the Fourier frequency of the
perturbation and $\Delta\equiv r(r-2M)$.  The boundary conditions dictate that
we should have outgoing waves at infinity and ingoing waves at the BH horizon:
\be
X_{lm}=\left\{
\begin{array}{l}
X_{lm}^{\rm in}e^{-i\omega r_*}\,,\quad r_*\to-\infty\,,\\
X_{lm}^{\rm out}e^{i\omega r_*}\,,\quad r_*\to+\infty\,.\\
\end{array}
\right.
\ee
The source term $S_{lm}$ in the Sasaki-Nakamura equation (\ref{SNeq}) is
determined by the point-particle trajectory.  Without loss of generality we
assume the trajectory to be an ``equatorial'' ($\theta=\pi/2$) timelike
geodesic in the Schwarzschild background, parametrized by
\beq
R^2\dot{R}^2&=&R^2(E^2-1)-f(R)\,L_z^2+2MR\,,\\
R^2\dot{\phi}&=&L_z\,,\quad f(R)\dot{T}=E\,, \eeq
where $f(R)=1-2M/R$ and dots stand for derivatives with respect to proper time
$\tau$. As usual we denote by $L_z$ the orbital angular momentum of the
particle along the $z$-axis and by $E$ the particle's energy at infinity per
unit mass $\mu$, so $E=1$ corresponds to an infall from rest.  We remind the
reader that the impact parameter $b$ is related to $L_z$ by $b\equiv
L_z/\sqrt{E^2-1}$. The geodesic equations can be integrated numerically for
chosen values of $E$ and $L_z$.  For $L_z>L_{\rm crit}=L_{\rm crit}(E)$ (see
section~\ref{sec:Lcrit}) the particle does not plunge, but rather scatters to
infinity.  We restrict our discussion to the case $L<L_{\rm crit}$. Examples
of plunging orbits for $E=3$ and different angular momenta (i.e., different
impact parameters) are shown in Fig.~\ref{traj}.

Given a particle trajectory, the numerical calculation of the gravitational
radiation emitted by the system involves the solution of Eq.~(\ref{SNeq}),
which can be obtained by the standard Green's function technique (more details
on the formalism can be found in Appendix \ref{sec:SNformalism}). Omitting for
simplicity the indices $(l\,,m)$, we first define two independent solutions
$X_{\rm in}^{(0)}\,,X_{\rm out}^{(0)}$ with boundary conditions

\begin{subequations}
\beq
\label{SNsol1}
X_{\rm in}^{(0)}&=&\left\{
\begin{array}{l}
e^{-i\omega r_*}\,,\quad r_*\to-\infty\,,\\
A_{\rm out}e^{i\omega r_*}+A_{\rm in}e^{-i\omega r_*}\,,\quad r_*\to+\infty\,.\\
\end{array}
\right.
\\
\label{SNsol2}
X_{\rm out}^{(0)}&=&\left\{
\begin{array}{l}
B_{\rm out}e^{i\omega r_*}+B_{\rm in}e^{-i\omega r_*}\,,\quad r_*\to-\infty\,.\\
e^{i\omega r_*}\,,\quad r_*\to+\infty\,,\\
\end{array}
\right.
\eeq
\end{subequations}

Then the solution of the inhomogeneous Sasaki-Nakamura equation (\ref{SNeq})
is given by
\be
X_{lm}=\frac{1}{W}
\left[
X_{\rm in}^{(0)}  \int_{r_*}^\infty S_{lm} X_{\rm out}^{(0)}dr_*+
X_{\rm out}^{(0)} \int_{-\infty}^{r_*} S_{lm} X_{\rm in}^{(0)}dr_*
\right]\,.
\ee
where $W\equiv 2i\omega A_{\rm in}$ is the Wronskian. Asymptotically for
$r_*\to\infty$ the amplitude of the wave function is
\be
\label{convol}
X_{lm}^{\rm out}=\frac{1}{W}\int_{-\infty}^{\infty} S_{lm}X_{\rm in}^{(0)}dr_*\,.
\ee
Then the radiated energy, angular momentum and linear momentum are given by
the multipolar sums
\beq 
\label{Erad}
E^{\rm rad}&=&
%\int_{0}^{\infty}d\omega\,16\omega^2\sum_{lm}|X_{lm}^{\rm out}|^2=
\int_{0}^{\infty}d\omega\,\sum_{lm}\frac{dE_{lm}}{d\omega}\,,
\\
\label{Jrad}
J^{\rm rad}&=&\int_{0}^{\infty}d\omega\,
\sum_{lm}\frac{m}{\omega}\frac{dE_{lm}}{d\omega}\,,\\
\label{PxPy}
P_x+iP_y&=&\int_{0}^{\infty}d\omega\,16\omega^2 
\sum_{lm}{\biggl[}q_{lm}X^{\rm out}_{l\,,m}\bar{X}^{\rm out}_{l\,,m+1}+\\
&&p_{lm}\left(X^{\rm out}_{l\,,m}\bar{X}^{\rm out}_{l+1\,,m+1}-
\bar{X}^{\rm out}_{l\,,-m}X^{\rm out}_{l+1\,,-m-1}\right)
{\biggr]}\,,
\nonumber
\eeq
with
\beq
\frac{dE_{lm}}{d\omega}&=& 16\omega^2|X_{lm}^{\rm out}|^2\,,
\\
p_{lm}&=&
\left[\frac{(l+3)(l-1)(l+m+2)(l+m+1)}{(2l+3)(2l+1)(l+1)^2}\right]^{1/2}\,,
\nonumber
\\
q_{lm}&=&-2\frac{\sqrt{(l-m)(l+m+1)}}{l(l+1)}\,.
\nonumber
\eeq
Note that the radiated momentum in the $z$--direction $P_z=0$ by symmetry.
Details on the derivation of the source term and asymptotic expansions of the
wave functions which are necessary to improve the numerical accuracy of the
Wronskian are given in Appendix \ref{sec:SNformalism}.

We integrate all differential equations in {\sc C++} using the adaptive
stepsize integrator {\sc StepperDopr5} \cite{NR}. Schematically, the
integration consists of the following steps: (i) integrate the first
independent solution of the homogeneous SN equation (\ref{SNeq}) with the
boundary conditions (\ref{SNsol1}) from $r_h = 2M(1+\delta r)$ outwards
(typically we choose $\delta r=10^{-4}$); (ii) integrate the second
independent solution of the homogeneous equation with boundary conditions
(\ref{SNsol2}) from $r_{\infty}=r_{\infty}^{(0)}/\omega$ inwards, where
typically we choose $r_{\infty}^{(0)} = 4\times 10^4$; (iii) compute the
Wronskian at the large but finite radius $r_{\infty}^{(0)}$, using
Eq.~(\ref{wronsk:Ain}) for increased accuracy; (iv) integrate the geodesics
with given orbital parameters and at the same time compute the source term
using Eqs.~(\ref{Wsource}) and (\ref{Weqns}); (v) output the ``in'' solution
$X_{\rm in}^{(0)}$ and the source term $S_{lm}$ on a grid of $n=1.6\times
10^5$ colocation points, and use a Gauss-Legendre spectral integrator \cite{NR}
to compute the convolution of the homogeneous solutions with the source term
entering the expression for the outgoing amplitude $X_{lm}^{\rm out}$,
Eq.~(\ref{convol}); (vi) sum over multipoles to get the total radiated energy
(\ref{Erad}), the angular momentum (\ref{Jrad}) and the linear momentum in the
$x$-- and $y$--directions (\ref{PxPy}). 

In the remainder of this section we summarize the results obtained by this
procedure.

%%%%%%%%%%%%%%%%%%%%%%%%%%%%%%%%%%%%%%%%%%%%%%%%%%%%%%%%%%%%%%%%%%%%%%%%%%%%%%%
\subsection{Point-particle spectra}
%%%%%%%%%%%%%%%%%%%%%%%%%%%%%%%%%%%%%%%%%%%%%%%%%%%%%%%%%%%%%%%%%%%%%%%%%%%%%%%

We performed an extensive set of simulations, selecting seven values of the
normalized particle energies ($E=1,\,1.5,\,3,\,5,\,10,\,20,\,100$). For each
value of $E$ we considered nine different particle angular momenta
($L_z/L_{\rm crit}
=0,\,0.2,\,0.5,\,0.75,\,0.9,\,0.95,\,0.99,\,0.999,\,0.9999$), for a total of
63 different configurations. We also ran a few more cases to validate our code
against the results of Ref.~\cite{Oohara:1984bw}. Our results are in good
visual agreement with their plots.  For each of the 63 simulations we computed
all multipolar components of the radiation up to $l=l_{\rm max}=6$ for
$10^{-2}\leq \omega \leq 1.5$ in steps of $\delta \omega =10^{-2}$; in a few
selected cases we ran the code up to $l=l_{\rm max}=8$ to check the
convergence of the results against truncation of the multipolar sum. We also
verified that for head-on collisions with $E>1$ our results are in agreement
(within three decimal places) with Table II of Ref.~\cite{Lousto:1996sx}.

\begin{figure*}[htb]
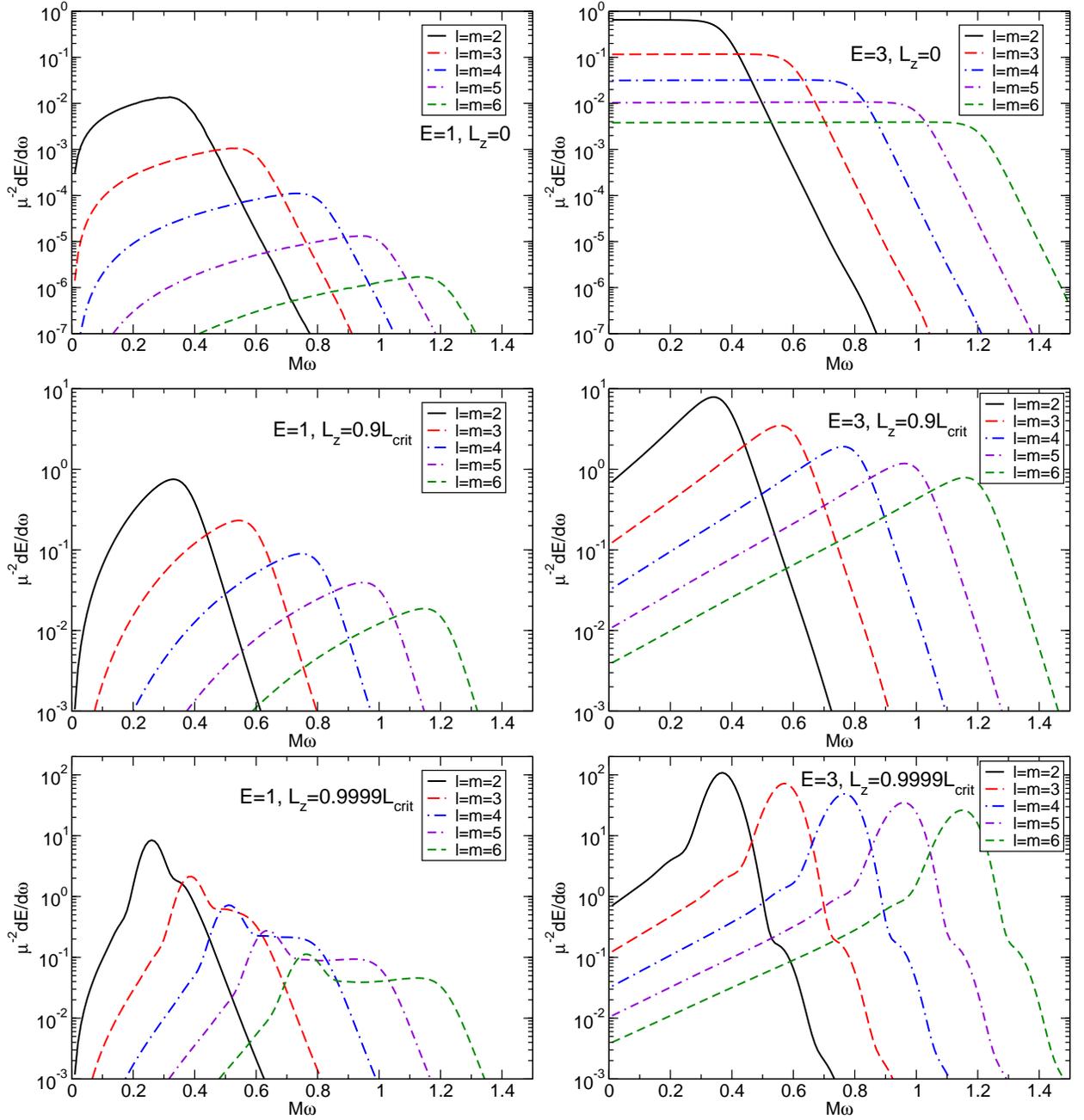

\begin{center}
\begin{tabular}{cc}
\includegraphics[scale=0.33,clip=true]{Fig10a.eps} &
\includegraphics[scale=0.33,clip=true]{Fig10b.eps} \\
\includegraphics[scale=0.33,clip=true]{Fig10c.eps} &
\includegraphics[scale=0.33,clip=true]{Fig10d.eps} \\
\includegraphics[scale=0.33,clip=true]{Fig10e.eps} &
\includegraphics[scale=0.33,clip=true]{Fig10f.eps} \\
\end{tabular}
\end{center}
\caption{\label{E1L0} Spectra for $l=m=2,\dots,6$ for nonrelativistic infalls
  ($E=1$, left column) and kinetic energy-dominates infalls ($E=3$, right
  column). The top row refers to a radial infall, the middle (bottom) row to
  an infall with $L/L_{\rm crit}=0.9$ ($L/L_{\rm crit}=0.9999$).}
\end{figure*}

For each value of $l$, most of the radiation is typically emitted in the $l=m$
component. In Fig.~\ref{E1L0} we show the $l=m$ component of the energy
spectra for particles falling from rest ($E=1$) and for kinetic-energy
dominated infalls ($E=3$). In each of these two cases we plot the spectra for
three selected values of the angular momentum: $L_z/L_{\rm crit}=0$ (a head-on
infall), $L_z/L_{\rm crit}=0.9$ and $L_z/L_{\rm crit}=0.9999$ (a near-critical
infall).  The particle trajectories corresponding to these values of $L_z$ are
shown in Fig.~\ref{traj} for $E=3$.

Not surprisingly, ultrarelativistic infalls radiate much more energy for a
given value of $L_z/L_{\rm crit}$. It is also apparent from Fig.~\ref{E1L0}
that the energy output increases (and higher multipoles become relatively more
important) as $L_z/L_{\rm crit}$ grows at fixed particle energy $E$. For
kinetic-energy dominated infalls the energy spectrum approaches a nonzero
constant as $M\omega\to 0$.  The spectrum is flat for small frequencies in the
head-on limit
\cite{Cardoso:2002ay,Cardoso:2002yj,Cardoso:2002jr,Berti:2003si}, but the
slope of the spectrum for $M\omega \ll1$ is nonzero when the infall is
nonradial. In all cases the energy spectrum decays exponentially at
frequencies $\omega>\omega_{\rm QNM}^l$, where $\omega_{\rm QNM}^l$ is the
fundamental Schwarzschild QNM frequency for the given multipole
\cite{Berti:2005ys,Berti:2009kk}.

\begin{figure*}[htb]
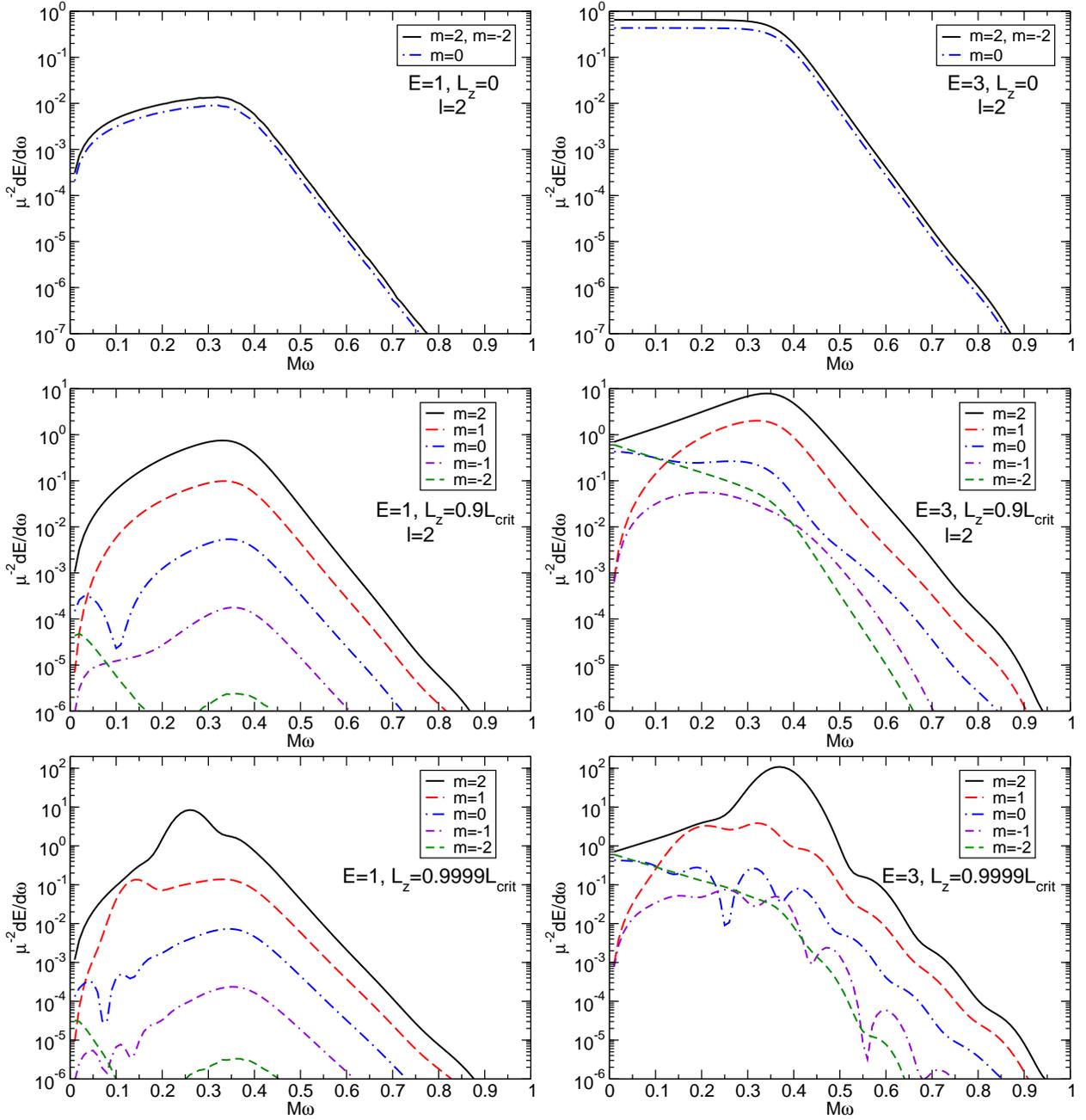

\begin{center}
\begin{tabular}{cc}
\includegraphics[scale=0.33,clip=true]{Fig11a.eps} &
\includegraphics[scale=0.33,clip=true]{Fig11b.eps} \\
\includegraphics[scale=0.33,clip=true]{Fig11c.eps} &
\includegraphics[scale=0.33,clip=true]{Fig11d.eps} \\
\includegraphics[scale=0.33,clip=true]{Fig11e.eps} &
\includegraphics[scale=0.33,clip=true]{Fig11f.eps} \\
\end{tabular}
\end{center}
\caption{\label{E1L0l2} Spectra for $l=2$ and $|m|\leq l$ for nonrelativistic
  infalls ($E=1$, left column) and kinetic energy-dominates infalls ($E=3$,
  right column). The top row refers to a radial infall, the bottom row to an
  infall with $L/L_{\rm crit}=0.9$.}
\end{figure*}

It is instructive to look at the multipolar structure of the radiation as we
vary $m$ for fixed $l$. Figure \ref{E1L0l2} shows spectra for the dominant
quadrupolar component ($l=2$) and all allowed values of $m$ ($|m|\leq l$) for
nonrelativistic infalls ($E=1$) and kinetic energy-dominates infalls ($E=3$)
with $L_z/L_{\rm crit}=0,\,0.9,\,0.9999$. In the head-on case the $(l+m)$--odd
components vanish, and components with the same $|m|$ are exactly equal to
each other because of symmetry. When $L_z\neq 0$ the particle motion breaks
this degeneracy: the odd--$m$
components are no longer zero, but they are still suppressed in the ZFL.
The $l=m$ ($l=-m$) components of the spectrum emerge from the ZFL with a
positive (negative) slope, respectively.  For large energies this slope is
independent of the particle energy and proportional to the impact parameter,
in qualitative agreement with the predictions of the toy model of
section~\ref{zflgrazing}.

For intermediate frequencies, the ZFL model of section~\ref{zflgrazing} would
predict a resonance in the spectrum at the rotational frequency of the rod,
and a $1/(\omega-\omega_0)$ dependence near resonance. The perturbative
spectra (as well as the NR spectra shown in section~\ref{sec:comp} below) are
instead characterized by an exponential decay for frequencies larger than the
fundamental QNM frequency of the given multipole.  This difference can be
attributed to the fundamentally different nature of the final state. In the
toy model the final state consists of two particles attached to a massless
rod, with rotation frequency determined by angular momentum conservation (and
an ad-hoc functional form of the stresses). In NR simulation and
point-particle infalls, the post-plunge dynamics is dominated by the QNMs of
the final BH.

\begin{table}[hbt]
\centering \caption{The numerically computed ZFL of the spectrum for different
  values of $E$ is compared to the analytical prediction from a multipolar
  decomposition of the ZFL formula (\ref{b0}), explained in Appendix
  \ref{app:multipoles}. The agreement is remarkable.} \vskip 12pt
\begin{tabular}{@{}ccccc@{}}
\hline \hline
&\multicolumn{4}{c}{$\frac{1}{(\mu E)^2}dE^{\rm rad}_{20}/d\omega |_{\omega=0}$}\\ \hline
$L_z/L_{\rm crit}$&$E=1.5$  &$E=3$   &$E=10$ &$E=100$\\
\hline \hline
$0.000$              &0.0160  &0.0481   &0.0644  &0.0663  \\
$0.500$              &0.0160  &0.0480   &0.0643  &0.0662  \\
$0.750$              &0.0159  &0.0480   &0.0642  &0.0662  \\
$0.900$              &0.0159  &0.0480   &0.0642  &0.0661  \\
$0.950$              &0.0159  &0.0479   &0.0641  &0.0661  \\
$0.990$              &0.0159  &0.0479   &0.0641  &0.0661  \\
$0.999$             &0.0159  &0.0479   &0.0641  &0.0661  \\
ZFL:                &0.0158  &0.0481   &0.0644  &0.0663  \\
\hline \hline
\end{tabular}
\label{tab:zfl}
\end{table}

The $m$--even components all tend to the ZFL as $M\omega\to 0$, while the
$m$--odd components are suppressed in the same limit. The dominant ($l=m$)
components of the spectra have a maximum corresponding to the QNM frequency
and decay (roughly) exponentially for $\omega>\omega_{\rm QNM}^l$. In
agreement with the toy model discussed in section~\ref{zflgrazing}, our
perturbative results indicate that the ZFL of the spectra depends very weakly
on the impact parameter, which (we recall again) is related to $L_z$ by
Eq.~(\ref{bcrit-eq}). This is shown very clearly in Table \ref{tab:zfl}.

A close inspection of the right panels of Fig.~\ref{E1L0l2} shows that the ZFL
of the spectrum for $m=0$ is not the same as for $m=\pm 2$. In fact, since the
ZFL is almost independent of $L_z$, one can use the head-on results to predict
the relative ratio between modes with $m=0$ and $|m|=2$. In the head-on case,
in a frame where the $z$--axis is aligned with the collision axis, only $m=0$
modes would contribute to the radiation. Once we rotate this ``natural''
coordinate system to the reference frame used in this paper, where a radial
infall occurs along the $x$--axis (see Fig.~\ref{traj}), for small $L_z$ the
spectrum for (say) $m=2$ will be related to the spectrum for $m=0$ by
\cite{Gualtieri:2008ux}:
\be
\lim_{\omega \to 0}{\frac{dE_{22}/d\omega}{dE_{20}/d\omega}}=\frac{3}{2}\,.
\ee
Our results for both equal-mass (see Fig.~\ref{fig:spceNUMR} below) and
extreme-mass ratio head-on collisions (Fig.~\ref{E1L0l2}) are in very good
agreement with this prediction.

\begin{figure}[htb]
\begin{center}
\includegraphics[scale=0.33,clip=true]{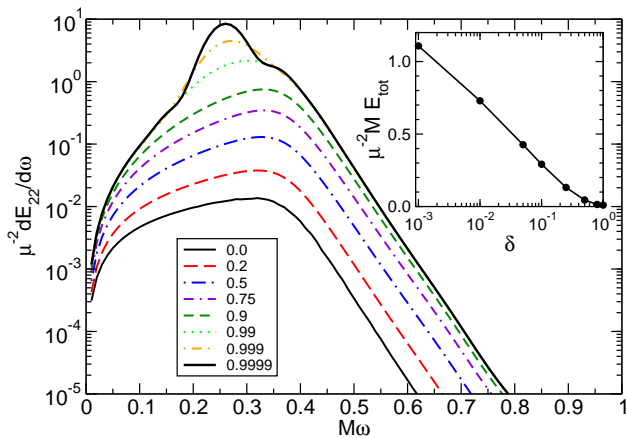}
\end{center}
\caption{Infall from rest: spectra for $l=m=2$ as $L_z/L_{\rm crit}\to 1$. In
  the inset: logarithmic divergence of the total radiated energy in the same
  limit.\label{fig:bump}}
\end{figure}

The energy spectra with $L_z/L_{\rm crit}=0.9,\,0.9999$ display secondary
peaks which are related to the orbital motion of the particle. Radiation of
orbital nature becomes more and more important as $L_z$ grows. Indeed, the
nature of the spectra changes quite significantly as $L_z/L_{\rm crit}\to 1$.
In Fig.~\ref{fig:bump} we show the $l=m=2$ component of the spectrum for
infalls from rest as we fine-tune the angular momentum to the critical value
for a plunge: a very distinctive ``bump'' appears at a frequency which is
slightly {\it lower} than the QNM frequency, significantly enhancing the
radiated energy. The location of this ``bump'' corresponds to (twice) the
orbital frequency of the particle at the marginally bound orbit, i.e.
$\omega=2\Omega_{\rm mb}=(4M)^{-1}$. 
For infalls from rest, a particle with $L=L_{\rm crit}$ will orbit around the
marginally bound circular geodesic.
The reason for this local maximum in the
spectrum is that when $L_z \to L_{\rm crit}$ the particle orbits a large
number of times close to the circular marginally bound orbit at $r=4M$.

Gravitational radiation significantly affects near-critical geodesics for
purely kinematical reasons.  As $L_z/L_{\rm crit}\to 1$ the particle circles
an infinite number of times around the marginally bound circular orbit with
radius $r=r_{\rm mb}$, taking an infinite amount of proper time to reach the
horizon.  The proximity of the orbit to criticality is conveniently described
by a small dimensionless parameter
\be
\delta\equiv 1-\frac{L_z}{L_{\rm crit}}\,. 
\label{deltaLcrit}
\ee
As $\delta\to 0$ the particle hovers at a circular geodesic $N\simeq
-\left(\Omega_{\rm crit}\dot{t}_{\rm crit}\,\log{(k\delta)}\right)/\left(\pi
\sqrt{2V_{\rm eff}''}\right)$ times before plunging \cite{Berti:2009bk}.  In
our case $V_{\rm eff}$ denotes the Schwarzschild effective radial potential,
so that $r^5V_{\rm eff}''=-24ML^2+6rL^2-4Mr^2$; the angular velocity
$\Omega\equiv d\phi/dt$, dots stand for derivatives with respect to proper
time and $k$ is a constant. All quantities are evaluated at the critical
circular geodesic with $L=L_{\rm crit}$, and radius $r=r_{\rm crit}$.
When $E=1$ this circular geodesic corresponds to the marginally bound orbit,
located at $r_{\rm mb}=4M$, $M\Omega_{\rm mb}=8^{-1}$ and
\be\label{NE1}
N\sim -\frac{1}{\pi\sqrt{2}}\log\delta\,.
\ee
In the ultrarelativistic limit $E\to \infty$ the critical circular geodesic is
located at the light ring $r_{\rm crit}=3M$, the corresponding orbital
frequency $M\Omega_{\rm crit}=(3\sqrt{3})^{-1}$ and
\be
N\sim -\frac{1}{2\pi}\log\left(2\delta\right)\,.
\ee 

The orbital frequency at the light ring is intimately related with the eikonal
(long-wavelength) approximation of the fundamental QNM frequency of a BH
\cite{pressringdown,mashhoon,Berti:2005eb,Cardoso:2008bp}. This implies that
ultrarelativistic infalls with a near-critical impact parameter are in a sense
the most ``natural'' and efficient process to resonantly excite the dynamics
of a BH. The proper oscillation modes of a Schwarzschild BH cannot be excited
by particles on {\it stable} circular orbits (i.e. particles with orbital
radii $r>6M$ in Schwarzschild coordinates), but near-critical
ultrarelativistic infalls are such that the orbital ``bump'' visible in Figure
\ref{fig:bump} moves just slightly to the right to overlap with the ``knee''
due to quasinormal ringing. So ultrarelativistic infalls have {\it just the
  right orbital frequency to excite BH oscillations}. Resonant gravitational
wave scattering explains the huge increase in radiated energy that can be
observed in the bottom right panels of Figs.~\ref{E1L0} and \ref{E1L0l2} (see
also \cite{Gualtieri:2001cm,Pons:2001xs,Pani:2010em}).  Since QNMs are
essentially perturbations of circular orbits at the light ring, we can expect
that these conclusions will still apply to Kerr and even Kerr-Newman BHs
\cite{mashhoon,Berti:2005eb}.

For near-critical orbits, the bottom panels of Fig.~\ref{E1L0l2} show that the
peak in the $(l=2,m=2)$ component is located at twice the frequency
corresponding to the peak in the $(l=2,m=1)$ component. The source term for a
particle in circular orbit with frequency $\Omega$ typically contains a term
proportional to $\delta(\omega-m\Omega)$; this lends further support to the
``orbital'' nature of the radiation enhancement. In fact, the inset of
Fig.~\ref{fig:bump} shows that the {\it total} radiated energy in the limit
$L_z\to L_{\rm crit}$ scales logarithmically with $\delta$, and hence linearly
with the number of orbits $N$, as expected for orbital radiation in the
nonresonant case. In fact, in the next subsection we will show that our
numerics are in {\it quantitative} agreement with the energy output of a
particle in circular orbit at the marginally bound geodesic.

%%%%%%%%%%%%%%%%%%%%%%%%%%%%%%%%%%%%%%%%%%%%%%%%%%%%%%%%%%%%%%%%%%%%%%%%%%%%%%%
\subsection{Energy distribution\label{sec:totenergy}}
%%%%%%%%%%%%%%%%%%%%%%%%%%%%%%%%%%%%%%%%%%%%%%%%%%%%%%%%%%%%%%%%%%%%%%%%%%%%%%%

We now turn our attention to the total integrated energy. To simplify the
discussion we start by revisiting and extending the analysis of infalls from
rest ($E=1$) first carried out by Oohara and Nakamura
\cite{Oohara:1984bw}. This reanalysis is useful both as a code check and to
stress some important characteristics of the radiation in the near-critical
limit $L_z/L_{\rm crit}\to 1$. Then we generalize our findings to infalls with
arbitrary energy.

\begin{figure*}[htb]
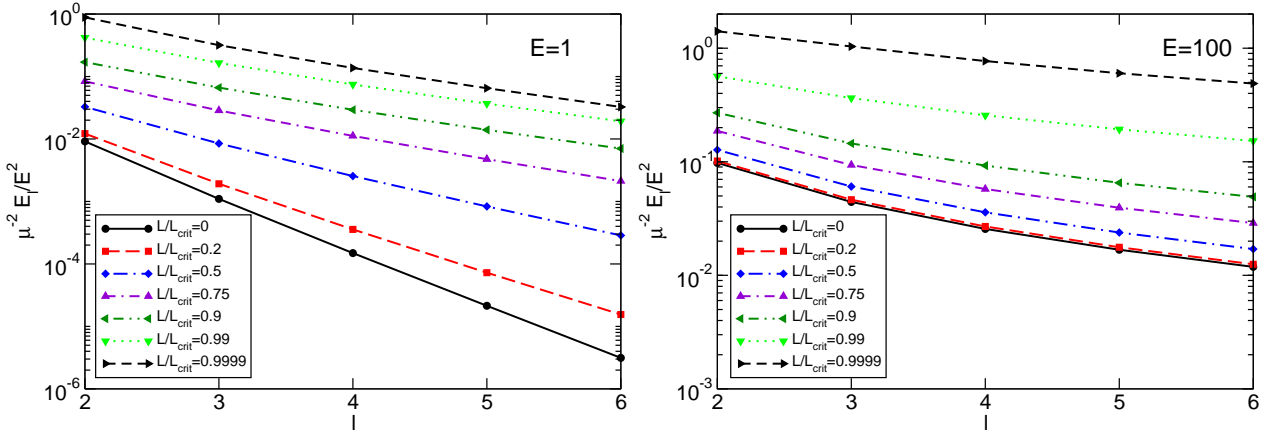

\begin{center}
\begin{tabular}{ccc}
\includegraphics[scale=0.33,clip=true]{Fig13a.eps} &
\includegraphics[scale=0.33,clip=true]{Fig13b.eps} \\
\end{tabular}
\end{center}
\caption{\label{El} Multipolar energy distribution for $E_l/E^2$ for $E=1$
  (left) and $E=100$ (right). For large $E$ the radiated energy scales like
  $E^2$, not like $\mu^2$. Higher multipoles contribute (relatively) more when
  $E$ is large. The scaling with $l$ is well approximated by an exponential
  for collisions from rest and by a power law for relativistic collisions.}
\end{figure*}
%

%%%%%%%%%%%%%%%%%%%%%%%%%%%%%%%%%%%%%%%%%%%%%%%%%%%%%%%%%%%%%%%%%%%%%%%%%%%%%%%
\subsubsection{Infall from rest ($E=1$)\label{sec:energyrest}}
%%%%%%%%%%%%%%%%%%%%%%%%%%%%%%%%%%%%%%%%%%%%%%%%%%%%%%%%%%%%%%%%%%%%%%%%%%%%%%%

Any numerical calculation is necessarily limited to a finite number of
multipoles. In this paper we computed multipolar components of the radiation
up to $l=l_{\rm max}=6$, but to get the total radiated energy, angular
momentum and linear momentum we should in principle compute the infinite sums
of Eqs.~(\ref{Erad}), (\ref{Jrad}) and (\ref{PxPy}). This requires fitting the
numerical results for $l\leq l_{\rm max}$ by some analytical formula, and
extrapolating this formula to estimate the contribution from multipoles with
$l>l_{\rm max}$. For infalls from rest, our results are consistent with the
functional dependence proposed by Oohara and Nakamura \cite{Oohara:1984bw}:
\be
\label{Elfit} 
\frac{M}{\mu^2}E^{\rm rad}_l=a_Ee^{-b_E\,l}\,.
\ee
\begin{table}[hbt]
\centering \caption{Fitting coefficients in Eqs.~(\ref{Elfit}) and (\ref{Jlfit}). For each value
  of $L_z/L_{\rm crit}$ the first line refers to a fit including all
  multipoles, the second line to a fit dropping the $l=2$ multipole. A number
  such as 1.042(-2) means $1.042\times 10^{-2}$.} \vskip 12pt
\begin{tabular}{@{}ccccccc@{}}
\hline \hline
$L_z/L_{\rm crit}$& $a_E$ & $b_E$ & $M/\mu^2 E^{\rm rad}$&$a_J$ & $b_J$ & $1/\mu^2 J^{\rm rad}$\\
\hline \hline
$0.000$          & 0.45 & 1.99 &1.04(-2)             &0.00  &0.00  & 0.00 \\
                  & 0.38 & 1.95 &1.04(-2)             &0.00  &0.00  & 0.00   \\
$0.200$          & 0.30 & 1.66 &1.46(-2)             &1.71  &1.66  & 8.20(-2)  \\
                  & 0.23 & 1.61 &1.46(-2)             &1.29  &1.61  & 8.21(-2)  \\
$0.500$          & 0.31 & 1.18 &4.50(-2)             &2.51  &1.24  & 3.22(-1) \\
                  & 0.24 & 1.13 &4.50(-2)             &1.75  &1.16  & 3.23(-1) \\
$0.750$          & 0.47 & 0.92 &1.32(-1)             &3.60  &0.96  & 8.98(-1) \\
                  & 0.37 & 0.87 &2.32(-1)             &2.58  &0.90  & 8.99(-1) \\
$0.900$          & 0.76 & 0.79 &2.92(-1)             &5.59  &0.83  & 1.94 \\
                  & 0.61 & 0.75 &2.93(-1)             &4.14  &0.77  & 1.95 \\
$0.950$          & 1.02 & 0.76 &4.27(-1)             &7.53  &0.80  & 2.85  \\
                  & 0.82 & 0.72 &4.28(-1)             &5.63  &0.75  & 2.86 \\
$0.990$          & 1.76 & 0.77 &7.30(-1)             &13.09  &0.80  & 4.99 \\
                  & 1.36 & 0.72 &7.32(-1)             &9.68  &0.74  & 5.01\\
\hline \hline
\end{tabular}
\label{tab:abON}
\end{table}

A fit of the data yields the coefficients listed in Table \ref{tab:abON}.  The
fitting coefficients listed in the first row are obtained by fitting all data
($2\leq l \leq 6$). The lowest multipole $l=2$ is usually an outlier in this
fit. As a rough check of the accuracy of the extrapolation, we repeat the fit
considering only numerical data with $3\leq l \leq 6$; this yields the
coefficients listed in the second row. The difference between the total
energies obtained by these two procedures can be seen as a very rough estimate
of the error involved in the extrapolation. It is quite clear from the table
that this error increases as $L_z/L_{\rm crit}\to 1$.  This is of course a
lower limit on the overall error in the computed energy, because it does not
take into account numerical errors in the data, systematic errors coming from
the (somewhat arbitrary) choice of the fitting function, and inaccuracies in
the fit itself. In any event, our fits are in good (but not perfect) agreement
with the entries in Table I of Oohara and Nakamura
\cite{Oohara:1984bw}. Oohara and Nakamura do not specify the range in $l$ used
for their fits, so it is hard to say if the slight difference in fitting
coefficients is due to small differences in the numerics or to the fitting
procedure itself.

\begin{figure}
\begin{center}
\begin{tabular}{c}
\includegraphics[scale=0.33,clip=true]{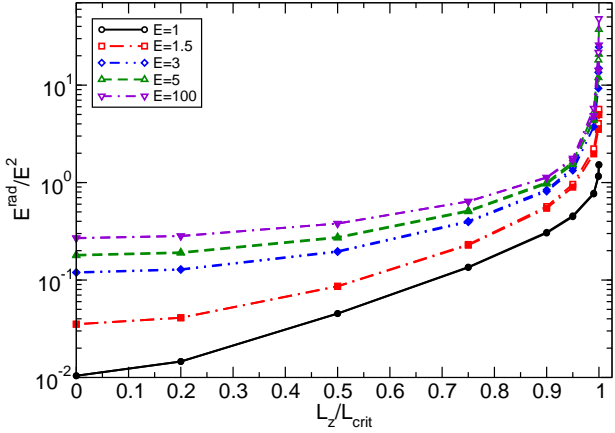}
\end{tabular}
\end{center}
\caption{\label{fig:Erad} Total energy radiated $M/(\mu E)^2 E^{\rm rad}$
  rescaled by the particle energy squared, as a function of $L_z/L_{\rm
    crit}$. In extrapolating for $l>6$, the fitting coefficients in
  Eqs.~(\ref{Elfit}) and (\ref{ENfitlargeE}) were obtained either including
  (empty symbols) or excluding (filled symbols) the $l=2$ data. The difference
  is only visible in the limit $L_z/L_{\rm crit}\to 1$.}
\end{figure}
A fit to our perturbative results close to the critical angular momentum for
capture yields
\be
\frac{M}{\mu^2}E^{\rm rad}=-0.0375-0.172\log\delta
\,.
\label{numcritE1}
\ee
where $\delta$ was defined in Eq.~(\ref{deltaLcrit}).  

The analysis of the previous section reveals that close to the critical value
of the angular momentum (i.e. as $\delta\to 0$) the total radiated energy
$E^{\rm rad}$ should scale as
\be 
\label{EN1}
\frac{M}{\mu^2}E^{\rm rad} \simeq k+E_{N=1}N\,, 
\ee
where $E_{N=1}$ is the energy radiated in one revolution close to the circular
marginally bound geodesic, and $k$ is an undetermined constant. Early
calculations by Detweiler \cite{Detweiler:1978ge,Smarr} (that we confirmed
using a {\sc Fortran} code discussed, for example, in
Ref.~\cite{Yunes:2008tw}) show that $\frac{M}{\mu^2}\,E_{N=1} \simeq 0.65$ for
circular orbits with $r_{\rm crit}=4M$ in Schwarzschild.
%For l=|m|=2 at r=4M I get 2*0.00851806=0.0170361: see Fig.1, Detweiler:1978ge
%
Using Eq.~(\ref{NE1}) one gets the independent estimate
\be 
\frac{M}{\mu^2}E^{\rm rad} \simeq k-0.15\log\delta\,,
\ee
which is in good agreement with the fit of Eq.~(\ref{numcritE1}).

%%%%%%%%%%%%%%%%%%%%%%%%%%%%%%%%%%%%%%%%%%%%%%%%%%%%%%%%%%%%%%%%%%%%%%%%%%%%%%%
\subsubsection{Ultrarelativistic infall}
%%%%%%%%%%%%%%%%%%%%%%%%%%%%%%%%%%%%%%%%%%%%%%%%%%%%%%%%%%%%%%%%%%%%%%%%%%%%%%%

For ultrarelativistic collisions the scaling with $l$ is not exponential, as
illustrated in Fig.~\ref{El}. In fact, we find that a power law of the form
\be 
\frac{ME^{\rm rad}_l}{(\mu\,E)^2}=\,c_El^{-d_E}\,, 
\label{ENfitlargeE}
\ee
describes well our numerical results, where $c_E$ and $d_E$ are constants. A
least-squares fit yields the values of $c_E$, $d_E$ and the total radiated
energy $E^{\rm rad}$ listed in Table \ref{tab:abON2}.

\begin{table*}[hbt]
\centering \caption{Fitting coefficients in Eq.~(\ref{ENfitlargeE}). For each
  value of $L_z/L_{\rm crit}$ the first line refers to a fit keeping all
  multipoles, the second line to a fit dropping the $l=2$ multipole.}
\vskip 12pt
\begin{tabular}{@{}ccccccccccccc@{}}
\hline \hline
&\multicolumn{3}{c}{$E=1.5$}&\multicolumn{3}{c}{$E=3$}&\multicolumn{3}{c}{$E=10$}&\multicolumn{3}{c}{$E=100$}\\ \hline
$L_z/L_{\rm crit}$ & $c_E$  & $d_E$&$\frac{M}{(\mu E)^2}E^{\rm rad}$  & $c_E$  & $d_E$&$\frac{M}{(\mu E)^2}E^{\rm rad}$  & $c_E$ & $d_E$ &$\frac{M}{(\mu E)^2}E^{\rm rad}$ &$c_E$ & $d_E$&$\frac{M}{(\mu E)^2}E^{\rm rad}$\\
\hline \hline
0.000    &1.75    &5.51 &0.035& 6.80(-1)   &3.10   &0.12 &3.99(-1)   &2.06   &0.24 &3.67(-1)   &1.92   &0.27\\
         &6.82    &6.39 &0.035& 1.11       &3.41   &0.12 &4.16(-1)   &2.08   &0.24 &3.56(-1)   &1.90   &0.27\\
0.200    &1.26    &4.96 &0.041& 6.71(-1)   &3.01   &0.13 &4.09(-1)   &2.04   &0.25 &3.78(-1)   &1.90   &0.28\\
         &3.74    &5.66 &0.041& 1.06       &3.31   &0.13 &4.28(-1)   &2.07   &0.25 &3.69(-1)   &1.89   &0.28\\
0.500    &7.24(-1)&3.50 &0.086& 6.61(-1)   &2.60   &0.20 &4.86(-1)   &1.93   &0.35 &4.52(-1)   &1.83   &0.38\\
         &1.26    &3.86 &0.086& 9.08(-1)   &2.81   &0.19 &5.11(-1)   &1.97   &0.34 &4.53(-1)   &1.83   &0.38\\
0.750    &7.38(-1)&2.56 &0.23& 7.50(-1)    &2.13   &0.40 &6.55(-1)   &1.77   &0.61 &6.06(-1)   &1.70   &0.64\\
         &1.01    &2.76 &0.23& 8.90(-1)    &2.24   &0.39 &6.82(-1)   &1.79   &0.60 &6.06(-1)   &1.70   &0.64\\
0.900    &0.91    &2.02 &0.57& 8.86(-1)    &1.76   &0.84 &8.42(-1)   &1.58   &1.14 &7.95(-1)   &1.55   &1.13\\
         &1.16    &2.17 &0.54& 9.94(-1)    &1.83   &0.81 &8.85(-1)   &1.61   &1.11 &8.08(-1)   &1.56   &1.12\\
0.950    &1.10    &1.80 &0.96& 1.00        &1.56   &1.41 &9.75(-1)   &1.45   &1.80 &9.33(-1)   &1.43   &1.77\\
         &1.41    &1.96 &0.90& 1.12        &1.63   &1.33 &1.04       &1.49   &1.71 &9.75(-1)   &1.46   &1.71\\
0.990    &1.81    &1.62 &2.22& 1.41        &1.27   &4.48 &1.36       &1.20   &5.76 &1.32       &1.20   &5.77\\
         &2.42    &1.81 &1.97& 1.62        &1.36   &3.74 &1.51       &1.27   &4.78 &1.45       &1.26   &4.86\\
\hline \hline
\end{tabular}
\label{tab:abON2}
\end{table*}

For $L_z=0$ and large $E$ we get $E^{\rm rad}/(\mu E)^2=0.26$, in agreement
with the results of Refs.~\cite{Cardoso:2002ay,Berti:2003si}. Close to the
critical angular momentum ($L_z\to L_{\rm crit}$) the fit yields $E_l \sim
1/l$.  As explained in section~\ref{sec:totenergy}, in the limit $E\to \infty$
the radiation is dominated by energy emitted at the light ring (the marginally
bound circular geodesic located at $r=3M$). Our fit is perfectly consistent
with the classic study of synchrotron radiation by Davis {\it et al.}
\cite{Davis:1972dm}, who found precisely a $1/l$ dependence for the multipolar
dependence of the radiation emitted by a particle orbiting at the circular
null geodesic.

For $L_z/L_{\rm crit}<0.95$ we get the following fits for the total radiated
energy as a function of the angular momentum (see Fig.~\ref{fig:Erad}):
\beq
\frac{ME^{\rm rad}}{(\mu\,E)^2}&=&
0.0685\exp{\left[
3.241 \left(\frac{L_z}{L_{\rm crit}}\right)^4
\right]}\,,(E=1.5)\,,\label{erade1p5}
\nonumber\\
\frac{ME^{\rm rad}}{(\mu\,E)^2}&=&
0.145\exp{\left[
2.778 \left(\frac{L_z}{L_{\rm crit}}\right)^4
\right]}\,,(E=3)\,,\label{erade3}\\
\frac{ME^{\rm rad}}{(\mu\,E)^2}&=&
0.294\exp{\left[
2.176 \left(\frac{L_z}{L_{\rm crit}}\right)^4
\right]}\,,(E=100)\,.\nonumber
\eeq
These results should be compared with the fit of the energy computed from NR
simulations of equal-mass BHs \cite{Sperhake:2009jz}:
\be
\frac{E^{\rm rad}}{M_{\rm ADM}^2}=
0.0195\exp{\left[
2.632 \left(\frac{L_z}{L_{\rm crit}}\right)^4
\right]}\,,\quad(E=1.5)\,.
\label{eradnr}\ee
In the fit to NR results (which have a numerical error of about $5\%$) the
radiated energy is normalized to $M_{\rm ADM}$ and we estimate $L_z/L_{\rm
  crit}$ to be given by $b/b_{\rm crit}$
\cite{Sperhake:2009jz}. Unfortunately, close to $L_{\rm crit}$ it is extremely
difficult to get accurate estimates for $E^{\rm rad}$, because higher
multipoles make an important contribution to the total radiation.

%%%%%%%%%%%%%%%%%%%%%%%%%%%%%%%%%%%%%%%%%%%%%%%%%%%%%%%%%%%%%%%%%%%%%%%%%%%%%%%%
\subsection{Angular momentum}
%%%%%%%%%%%%%%%%%%%%%%%%%%%%%%%%%%%%%%%%%%%%%%%%%%%%%%%%%%%%%%%%%%%%%%%%%%%%%%%%

%
\begin{figure}
\begin{center}
\includegraphics[scale=0.33,clip=true]{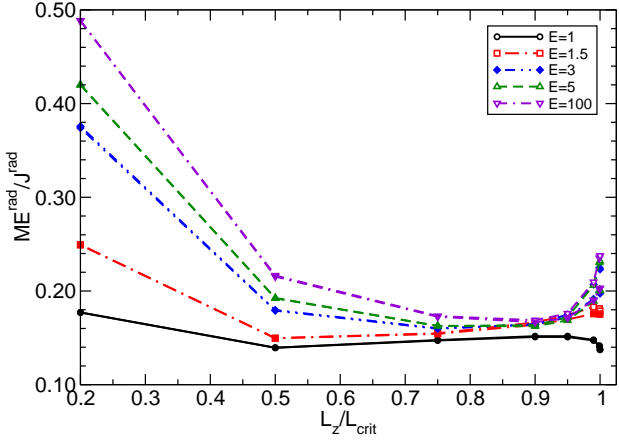}
\end{center}
\caption{\label{fig:JE} Ratio $M E^{\rm rad}/J{\rm rad}$ for different values
  of $E$. The meaning of filled and empty symbols is the same as in
  Fig.~\ref{fig:Erad}. Oohara and Nakamura \cite{Oohara:1984bw} find that this
  ratio is $\simeq 0.15\pm 0.01$ for $1\lesssim L_z\lesssim 3.9$ when $E=1$
  and $L_{\rm crit}=4$, but we expect that $E^{\rm rad}/J^{\rm rad}\simeq
  (M\Omega_{\rm scat})$ as $L_z/L_{\rm crit}\to 1$ (see text).}
\end{figure}
\begin{figure}
\begin{center}
\includegraphics[scale=0.33,clip=true]{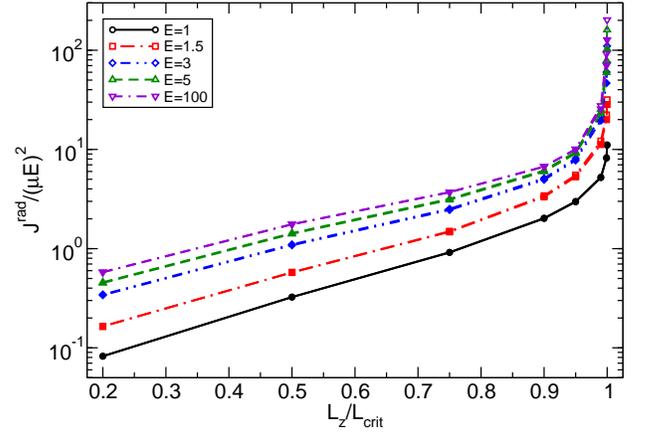}
\end{center}
\caption{\label{fig:Jrad} Total radiated angular momentum (obtained by
  extrapolation) as a function of $L_z/L_{\rm crit}$. The meaning of filled
  and empty symbols is the same as in Fig.~\ref{fig:Erad}.}
\end{figure}
\begin{figure*}[htb]
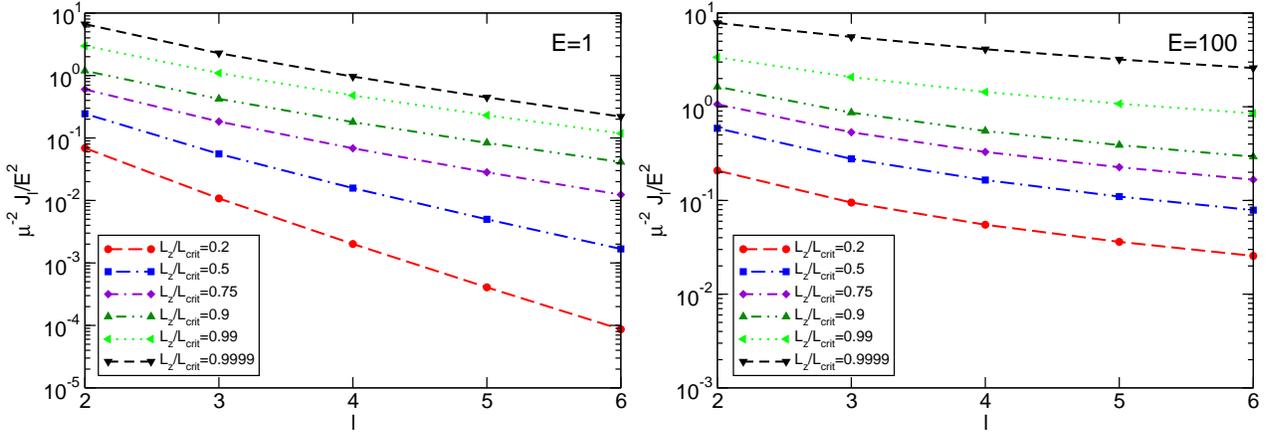

\begin{center}
\begin{tabular}{ccc}
\includegraphics[scale=0.33,clip=true]{Fig17a.eps}&
\includegraphics[scale=0.33,clip=true]{Fig17b.eps}\\
\end{tabular}
\end{center}
\caption{\label{fig:Jl} Multipolar components of the radiated angular momentum
  $J_l/E^2$ for $E=1$ (left) and $E=100$ (right). For large $E$ the radiated
  angular momentum scales like $(\mu E)^2$. Higher multipoles contribute
  (relatively) more when $E$ is large. The scaling with $l$ is logarithmic for
  collisions from rest, but it acquires corrections for relativistic
  collisions. By symmetry, no angular momentum is radiated when $L_z=0$.}
\end{figure*}
%

%%%%%%%%%%%%%%%%%%%%%%%%%%%%%%%%%%%%%%%%%%%%%%%%%%%%%%%%%%%%%%%%%%%%%%%%%%%%%%%
\subsubsection{Infall from rest ($E=1$)}
%%%%%%%%%%%%%%%%%%%%%%%%%%%%%%%%%%%%%%%%%%%%%%%%%%%%%%%%%%%%%%%%%%%%%%%%%%%%%%%

Let us consider the angular momentum carried by the radiation.  For slow
motion, the multipolar decomposition of the radiated angular momentum is
consistent with the exponential dependence
\be\label{Jlfit}
\frac{1}{\mu^2}J^{\rm rad}_l=a_Je^{-b_J\,l}\,,
\ee
where the coefficients $a_J,\,b_J$ are given in Table \ref{tab:abON}. 

As shown in Fig.~\ref{fig:JE}, our results for infalls from rest are
consistent with Oohara and Nakamura's \cite{Oohara:1984bw} suggestion that for
$0.25\lesssim L_z/L_{\rm crit}\lesssim0.99$ the radiated angular momentum
should obey the approximate relation
\be
\frac{M E^{\rm rad}}{J^{\rm rad}}\approx \left(0.15\pm 0.01\right)\,.
\label{approxejrad}
\ee
However, this approximate proportionality relation must break down as $L_z\to
L_{\rm crit}$, because for a particle in circular motion $E^{\rm rad}/J^{\rm
  rad}\simeq (M\Omega_{\rm scat})$ \cite{Detweiler:1978ge,Cutler:1994pb}. For
example, for $E=1$ we should find that $E^{\rm rad}/J^{\rm rad}\to 1/8=0.125$,
and as $E\to \infty$ we expect that $E^{\rm rad}/J^{\rm rad}\to
1/(3\sqrt{3})\simeq 0.19245$. Quite remarkably, and despite the sizeable
uncertainties in extrapolating our numerical results to get the total radiated
energy and angular momentum, Fig.~\ref{fig:JE} is consistent with these
predictions.

%%%%%%%%%%%%%%%%%%%%%%%%%%%%%%%%%%%%%%%%%%%%%%%%%%%%%%%%%%%%%%%%%%%%%%%%%%%%%%%
\subsubsection{Ultrarelativistic infall}
%%%%%%%%%%%%%%%%%%%%%%%%%%%%%%%%%%%%%%%%%%%%%%%%%%%%%%%%%%%%%%%%%%%%%%%%%%%%%%%

The total radiated angular momentum for generic infall energies and its
multipolar decomposition are shown in Figs.~\ref{fig:Jrad}-\ref{fig:Jl}.  For
collisions with generic impact parameter, we find that the angular momentum
emitted in a given multipole $l$ has a power-law dependence of the form
\be 
\frac{J^{\rm rad}_l}{(\mu\,E)^2}=\,c_Jl^{-d_J}\,, \label{JfitlargeE}
\ee
where $c_J$ and $d_J$ are constants. The values of $c_J$, $d_J$ and of the
total radiated angular momentum $J^{\rm rad}$ obtained by extrapolation are
listed in Table \ref{tab:Jfit}. The angular momentum emitted in each multipole
$l\leq l_{\rm max}$ is shown in Fig.~\ref{fig:Jl}. Using
Eq.~(\ref{JfitlargeE}) we can extrapolate to get the total radiated angular
momentum.  Fig.~\ref{fig:JE} shows that the ratio $J^{\rm rad}/(M E^{\rm
  rad})$ for kinetic-energy dominated infalls is only (roughly) constant in
the intermediate regime $0.5\lesssim L_z/L_{\rm crit}\lesssim 0.95$. The
following expressions provide good fits to the extrapolated values:
\beq
\frac{J^{\rm rad}}{(\mu\,E)^2}&=&
0.0142\left[
1-\exp{\left(3.170\frac{L_z}{L_{\rm crit}}\right)}
\right]^2\,,(E=1.5)\,,
\nonumber\\
\frac{J^{\rm rad}}{(\mu\,E)^2}&=&
0.0676\left[
1-\exp{\left(2.585\frac{L_z}{L_{\rm crit}}\right)}
\right]^2\,,(E=3)\,,
\nonumber\\
\frac{J^{\rm rad}}{(\mu\,E)^2}&=&
0.42\left[
1-\exp{\left(1.836\frac{L_z}{L_{\rm crit}}\right)}
\right]^2\,,(E=100)\,.
\nonumber\\
\eeq
These fits can be compared with the corresponding fit from NR simulations of
equal-mass BHs \cite{Sperhake:2009jz}, which yields (within a numerical error
of about $5\%$)
\be
\frac{J^{\rm rad}}{M_{\rm ADM}^2}=
0.0024\left[
1-\exp{\left(2.928\frac{L_z}{L_{\rm crit}}\right)}
\right]^2\,,\quad(E=1.5)\,.\label{jradnr}\\
\ee
\begin{table*}[hbt]
\centering \caption{Fitting coefficients in Eq.~(\ref{JfitlargeE}). For each
  value of $L_z/L_{\rm crit}$ the first line refers to a fit keeping all
  multipoles, the second line to a fit dropping the $l=2$ multipole.}
\vskip 12pt
\begin{tabular}{@{}ccccccccccccc@{}}
\hline \hline
&\multicolumn{3}{c}{$E=1.5$}&\multicolumn{3}{c}{$E=3$}&\multicolumn{3}{c}{$E=10$}&\multicolumn{3}{c}{$E=100$}\\ \hline
$L_z/L_{\rm crit}$   & $c_J$ & $d_J$ &$1/(\mu E)^2J^{\rm rad}$& $c_J$ & $d_J$ &$1/(\mu E)^2J^{\rm rad}$   & $c_J$ & $d_J$ &$1/(\mu E)^2J^{\rm rad}$ &$c_J$ & $d_J$&$1/(\mu E)^2J^{\rm rad}$\\
\hline \hline
0.000   &0.00   &0.00  &0.00  &0.00  &0.00  &0.00  &0.00   &0.00  &0.00  &0.00  &0.00  &0.00\\
        &0.00   &0.00  &0.00  &0.00  &0.00  &0.00  &0.00   &0.00  &0.00  &0.00  &0.00  &0.00\\
0.200   &3.79   &4.60  &0.16  &1.30  &2.71  &0.35  &0.82   &1.98  &0.54  &0.78  &1.91  &0.58\\
        &11.49  &5.32  &0.16  &1.89  &2.95  &0.34  &0.83   &1.99  &0.54  &0.76  &1.89  &0.58\\
0.500   &5.22   &3.58  &0.58  &3.11  &2.45  &1.10  &2.20   &1.89  &1.68  &2.09  &1.83  &1.75\\
        &9.62   &3.98  &0.57  &4.15  &2.64  &1.08  &2.24   &1.90  &1.67  &2.05  &1.82  &1.76\\
0.750   &5.55   &2.70  &1.50  &4.48  &2.09  &2.52  &3.62   &1.72  &3.65  &3.42  &1.69  &3.69\\
        &7.60   &2.90  &1.47  &5.40  &2.21  &2.45  &3.72   &1.74  &3.61  &3.38  &1.68  &3.71\\
0.900   &6.59   &2.15  &3.41  &5.76  &1.79  &5.11  &5.11   &1.58  &6.88  &4.84  &1.56  &6.70\\
        &8.09   &2.28  &3.32  &6.46  &1.87  &4.94  &5.25   &1.60  &6.77  &4.82  &1.56  &6.72\\
0.950   &7.77   &1.93  &5.51  &6.65  &1.63  &8.09  &6.11   &1.48  &10.40 &5.83  &1.47  &10.08\\
        &9.43   &2.06  &5.29  &7.30  &1.69  &7.75  &6.33   &1.50  &10.13 &5.89  &1.48  &9.98\\
0.990   &11.85  &1.71  &12.15 &9.02  &1.35  &21.65 &8.45   &1.26  &27.81 &8.14  &1.26  &27.54\\
        &15.02  &1.86  &11.22 &9.96  &1.42  &19.52 &8.98   &1.30  &25.35 &8.53  &1.29  &25.55\\
\hline \hline\end{tabular}\label{tab:Jfit}
\end{table*}
%

%%%%%%%%%%%%%%%%%%%%%%%%%%%%%%%%%%%%%%%%%%%%%%%%%%%%%%%%%%%%%%%%%%%%%%%%%%%%%%%%
\subsection{Linear momentum\label{sec:lm}}
%%%%%%%%%%%%%%%%%%%%%%%%%%%%%%%%%%%%%%%%%%%%%%%%%%%%%%%%%%%%%%%%%%%%%%%%%%%%%%%%

%
\begin{figure*}[htb]
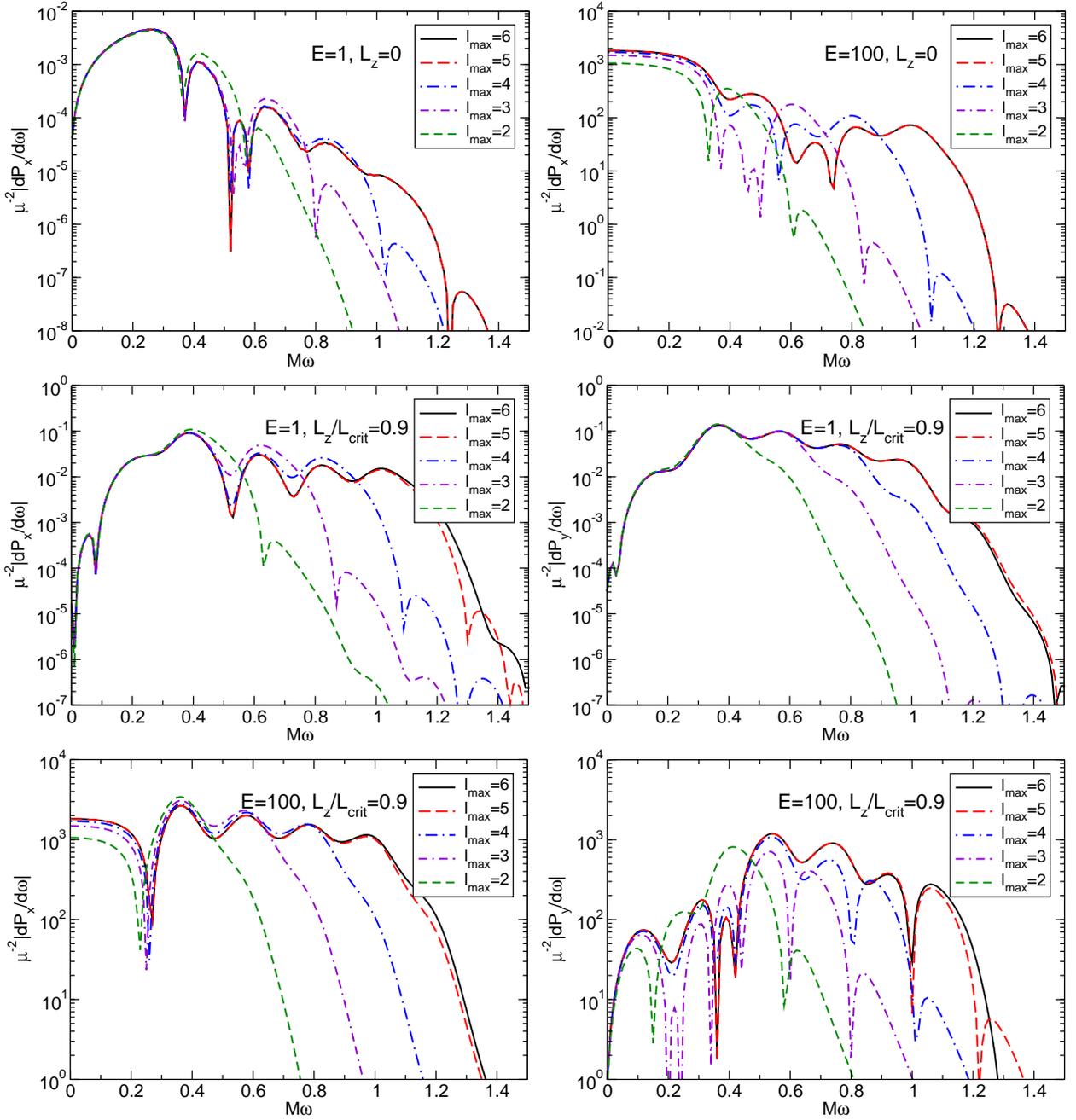

\begin{center}
\begin{tabular}{cc}
\includegraphics[scale=0.33,clip=true]{Fig18a.eps} &
\includegraphics[scale=0.33,clip=true]{Fig18b.eps} \\
\includegraphics[scale=0.33,clip=true]{Fig18c.eps} &
\includegraphics[scale=0.33,clip=true]{Fig18d.eps}\\
\includegraphics[scale=0.33,clip=true]{Fig18e.eps} &
\includegraphics[scale=0.33,clip=true]{Fig18f.eps}\\
\end{tabular}
\end{center}
\caption{\label{Pspectra} Spectra of the linear momentum radiated for infalls
  from rest ($E=1$) and ultrarelativistic infalls ($E=100$). Different lines
  refer to different truncation choices in the sum of Eq.~(\ref{PxPy}). In the
  head-on case (top row) the only nonzero component of the linear momentum is
  $P_x$; for $L/L_{\rm crit}=0.9$ (middle and bottom rows) both the $P_x$ and
  $P_y$ components are nonzero. To test convergence we sum Eq.~(\ref{PxPy}) up
  to different values of $l_{\rm max}$, as indicated in the legend.}
\end{figure*}

Representative spectra of the radiated linear momentum are shown in
Fig.~\ref{Pspectra}.  To illustrate how different multipoles contribute in
building up the total linear momentum, and also to ``visually'' test the
convergence of the sum in Eq.~(\ref{PxPy}), we plot in different linestyles
the spectra obtained by summing over the lowest $l_{\rm max}$ multipoles with
$l_{\rm max}=2\,,\dots\,,6$. Strictly speaking, if we sum Eq.~(\ref{PxPy}) up
to $l_{\rm max}=6$ the last term of the sum will be inconsistent, because it
involves amplitudes such as $X_{l+1,\,m+1}$, which we have not computed and we
set to zero for lack of a better alternative. Note that this ``multipolar
coupling'' does not occur in the calculation of the energy and angular
momentum. The inconsistent truncation means that we should only consider these
plots as representative of the real convergence properties of the sum for
$l_{\rm max}\leq 5$.  Furthermore, the spectrum obtained when we truncate at
$l_{\rm max}=6$ cannot be trusted in fitting numerical results to extrapolate
the sum to infinity.  For this reason (and also for the highly oscillatory
buildup of linear momentum which is evident from Fig.~\ref{Pspectra}) the
extrapolation of numerical results to get the total radiated linear momentum
is quite sensitive to the relatively low number of multipoles that we are
using.

%%%%%%%%%%%%%%%%%%%%%%%%%%%%%%%%%%%%%%%%%%%%%%%%%%%%%%%%%%%%%%%%%%%%%%%%%%%%%%%
\subsubsection{Infall from rest ($E=1$)}
%%%%%%%%%%%%%%%%%%%%%%%%%%%%%%%%%%%%%%%%%%%%%%%%%%%%%%%%%%%%%%%%%%%%%%%%%%%%%%%

Despite these caveats, in the case of low-energy collisions (and in particular
for infalls from rest) the total linear momentum converges reasonably fast
with $l$.  Summing over $2\leq l\leq 6$ is typically enough to warrant an
accuracy of $10\%$ or less in the total linear momentum.

Nevertheless we have used extrapolation, fitting our numerical results by an
exponential:
\be
P_{l_{\rm max}}=P^{\rm rad}-a_P e^{-b_P\,l_{\rm max}}\,,\label{E1Pfit}
\ee
The fitting coefficients and the total radiated momentum obtained by
extrapolation are listed in Table~\ref{tab:abONmomentum}.

\begin{table}[hbt]
\centering \caption{Fitting coefficients in Eq.~(\ref{E1Pfit}).  The first
  line refers to a fit including all multipoles, the second line to a fit
  dropping $l=2$.} \vskip 12pt
\begin{tabular}{@{}ccccccc@{}}
\hline \hline
$L_z/L_{\rm crit}$& $a_P$ & $b_P$ & $M/\mu^2 P^{\rm rad}$\\
\hline \hline
$0.000$          & 7.03(-3) &2.00 & 8.33(-4)          \\
                  & 7.98(-3) &2.05 & 8.33(-4)              \\
$0.200$          & 1.09(-2) &1.56 & 2.41(-3)              \\
                  & 9.99(-3) &1.53 & 2.42(-3)              \\
$0.500$          & 3.05(-2) &1.13 & 1.16(-2)             \\
                  & 2.58(-2) &1.06 & 1.16(-2)              \\
$0.750$          & 6.54(-2) &0.84 & 3.37(-2)             \\
                  & 5.55(-2) &0.78 & 3.39(-2)              \\
$0.900$          & 9.24(-2) &0.67 &5.75(-2)             \\
                  & 8.13(-2) &0.61 &5.81(-2)             \\
$0.950$          & 8.85(-2) &0.58 &6.20(-2)             \\
                  & 7.83(-2) &0.52 &6.30(-2)              \\
$0.990$          & 6.77(-2) &0.72 &4.96(-2)              \\
                  & 4.03(-2) &0.48 &5.16(-2)             \\
\hline \hline
\end{tabular}
\label{tab:abONmomentum}
\end{table}
%

%%%%%%%%%%%%%%%%%%%%%%%%%%%%%%%%%%%%%%%%%%%%%%%%%%%%%%%%%%%%%%%%%%%%%%%%%%%%%%%
\subsubsection{Ultrarelativistic infall}
%%%%%%%%%%%%%%%%%%%%%%%%%%%%%%%%%%%%%%%%%%%%%%%%%%%%%%%%%%%%%%%%%%%%%%%%%%%%%%%

Spectra of the linear momentum radiated by ultrarelativistic infalls are shown
in Fig.~\ref{Pspectra}.  For head-on collisions, in the limit of large boosts
the radiated momentum is well described by
\be
P^{\rm rad}= \left(0.11-\frac{0.33}{E^{1.37}}\right) \frac{(\mu\,E)^2}{M}\,,
\quad E \to \infty\,.
\ee
\begin{table}
\centering \caption{The ZFL of the radiated linear momentum for head-on
  collisions, compared with the analytical result (\ref{zflmomentum}).} \vskip
12pt
\begin{tabular}{@{}cccccc@{}}
\hline \hline
&\multicolumn{5}{c}{$\frac{1}{(\mu E)^2}dP^{\rm rad}/d\omega|_{\omega=0}$}\\ \hline
$L_z/L_{\rm crit}$ &$E=1.5$ &$E=3$   &$E=5$&$E=10$ &$E=100$\\
\hline \hline
$0.00$             &0.0182 &0.100  &0.151&0.187  &0.203 \\
ZFL:               &0.0180 &0.0993 &0.150&0.189  &0.212\\
\hline \hline
\end{tabular}
\label{tab:zflP}
\end{table}

For ultrarelativistic infalls the ZFL of the spectrum is nonvanishing and it
depends very weakly on the impact parameter (we observed a similar trend when
studying the ZFL of the spectrum of the radiated energy).  The ZFL is very
well approximated by Smarr's formula, Eq.~(\ref{zflmomentum}). This is shown
quantitatively in Table~\ref{tab:zflP}.

We tried to fit our numerical results for high energy and generic impact
parameters by a power law of the form $P_{l_{\rm max}}=P^{\rm rad}-c_P\,l_{\rm
  max}^{-d_P}$.  For the reasons explained above, the errors associated with
this extrapolation procedure are significant, and we decided not to present
the extrapolated values of $P^{\rm rad}$.

Qualitatively, in the ultrarelativistic case the total radiated linear
momentum increases only mildly with the impact parameter: we find a maximum
increase of a factor $\sim 3$ relatively to the head-on case.  This
observation can be used to give a rough estimate of the maximum recoil that
could result from gravitational-wave emission in the ultrarelativistic
collision of nonspinning BHs.  If we take $\mu\,E/M=1/10$ we get a recoil
velocity of order $v\sim 10^{-3}$ for head-on collisions, and a maximum of
around $v\sim 3\times 10^{-3}$ for grazing collisions. Restoring physical
units, this would correspond to recoils in the range $300-900\, {\rm km}/{\rm
  s}$. It will be interesting to verify these estimates by NR simulations of
ultrarelativistic, comparable-mass BH binaries.

%%%%%%%%%%%%%%%%%%%%%%%%%%%%%%%%%%%%%%%%%%%%%%%%%%%%%%%%%%%%%%%%%%%%%%%%%%%%%%%%
\section{Comparing different approaches\label{sec:comp}}
%%%%%%%%%%%%%%%%%%%%%%%%%%%%%%%%%%%%%%%%%%%%%%%%%%%%%%%%%%%%%%%%%%%%%%%%%%%%%%%%

As discussed in Ref.~\cite{Sperhake:2008ga} (see also
Fig.~\ref{fig:headonspec}), NR results for the energy spectra from the
high-energy, head-on collision of two BHs are in very good agreement with the
ZFL predictions. Calculations of the radiation from point particles falling
into nonrotating BHs are in quantitative agreement with ZFL predictions in the
extreme-mass ratio limit (see for instance Table~\ref{tab:zfl}), and in
qualitative agreement with NR calculations. In this section we present a more
extensive comparison between these different approaches.

%%%%%%%%%%%%%%%%%%%%%%%%%%%%%%%%%%%%%%%%%%%%%%%%%%%%%%%%%%%%%%%%%%%%%%%%%%%%%%%%
\subsection{Head-on collisions}
%%%%%%%%%%%%%%%%%%%%%%%%%%%%%%%%%%%%%%%%%%%%%%%%%%%%%%%%%%%%%%%%%%%%%%%%%%%%%%%%

For head-on collisions, the ZFL results of section~\ref{zflgrazing} are in
qualitative and quantitative agreement with point-particle calculations. In
particular, a multipolar decomposition of ZFL calculations yields a flat
energy spectrum for all multipolar components (see Appendix
\ref{app:multipoles}). Point-particle calculations and NR simulations suggest
that the cutoff frequency for each multipole can be chosen to be the lowest
QNM frequency for the given multipole. A multipolar analysis is therefore
important to introduce a more ``natural'' and appealing cutoff frequency in
the ZFL spectrum, in contrast with Smarr's original, qualitative suggestions
\cite{Smarr:1977fy}.

In the eikonal limit, the real part of the fundamental QNM frequency for the
$l$-th multipole $\omega^l_{\rm QNM}$ is related to the frequency
$M\Omega_c=(3\sqrt{3})^{-1}$ of unstable circular null geodesics
\cite{Cardoso:2008bp,Berti:2009kk}:
\be
\omega^l_{\rm QNM}=l\Omega_c\,, 
\ee
Combining this approximation (which is surprisingly good even for low values
of $l$) with ZFL calculations of the energy spectrum at low frequencies, we
can estimate the total radiated energy as follows:
\be
E^{\rm rad}\sim 
\sum_{l=2}^\infty \frac{dE_l}{d\omega}\Big| _{\omega =0}\times \omega^l_{\rm QNM}\,.
\ee
At high energies the ZFL assumes a particularly simple form. Using the
multipolar decomposition results from Appendix \ref{app:multipoles} we get,
for the equal-mass case:
\beq 
E^{\rm rad}&\simeq &\sum_{\rm even \,l} \frac{4 (E M)^2}{\pi}
\frac{(2l+1)(l-2)!}{(l+2)!}l\Omega_c \nonumber \\
&=&\frac{4\log2}{3\sqrt{3}\pi}EM
\simeq 0.1698 EM\,.
\eeq
Here we used $EM\Omega_c=(3\sqrt{3})^{-1}$, since the final BH has
(approximately) mass $EM$.  An extrapolation of results from recent NR
simulations \cite{Sperhake:2008ga} predicts $E^{\rm rad}\simeq (0.14\pm
0.03)M_{\rm ADM}$, in remarkably good agreement with this naive estimate.

As another example, let us consider the extreme mass-ratio case in the
high-energy limit. Using again results from Appendix \ref{app:multipoles} we
get
\beq
E^{\rm rad}&\simeq&\sum_l \frac{4 E^2
\mu^2}{\pi}\frac{(2l+1)(l-2)!}{(l+2)!}l\Omega_c \nonumber \\
&=& \frac{13}{9\sqrt{3}\pi}\frac{\mu^2 E^2}{M}
\simeq 0.265 \frac{\mu^2 E^2}{M}\,.
\eeq
The extrapolation of point-particle results to $E\to \infty$ (see
\cite{Cardoso:2002ay,Berti:2003si} and Table \ref{tab:abON2}) predicts $E^{\rm
  rad}\sim 0.262 \mu^2 E^2/M$ \cite{Cardoso:2002ay}, again in remarkable
agreement with our simple approximation.

%%%%%%%%%%%%%%%%%%%%%%%%%%%%%%%%%%%%%%%%%%%%%%%%%%%%%%%%%%%%%%%%%%%%%%%%%%%%%%%%
\subsection{Collisions with finite impact parameter}
%%%%%%%%%%%%%%%%%%%%%%%%%%%%%%%%%%%%%%%%%%%%%%%%%%%%%%%%%%%%%%%%%%%%%%%%%%%%%%%%

%
\begin{figure*}[htb]
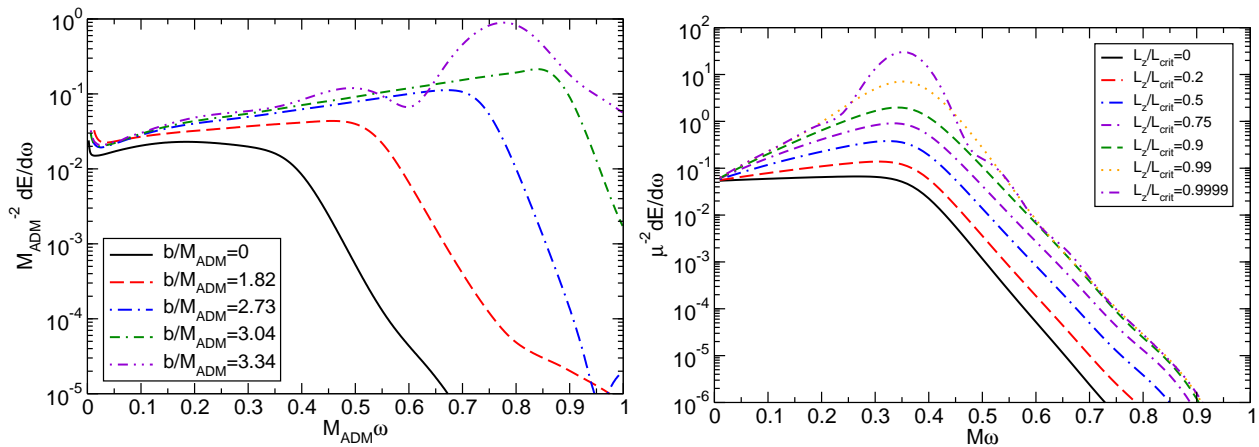

\begin{center}
\begin{tabular}{cc}
\includegraphics[scale=0.33,clip=true]{Fig19a.eps}&
\includegraphics[scale=0.33,clip=true]{Fig19b.eps}
\end{tabular}
\end{center}
\caption{\label{fig:spceNUMR} Left: $l=m=2$ component of the energy spectrum
  of the gravitational radiation emitted in the collision of two equal-mass
  BHs having speed $\beta=v/c\simeq 0.75$ in the center-of-mass frame
  \cite{Sperhake:2008ga,Sperhake:2009jz}. All quantities are normalized to the
  ADM mass of the system $M_{\rm ADM}$. Right: $l=m=2$ component of the energy
  spectrum of the gravitational radiation emitted by point particles falling
  into Schwarzschild BHs of mass $M$ with energy $E=1.5$.}
\end{figure*}

Unfortunately, an extension of our head-on estimates to the generic case of
collisions with finite impact parameter is not straightfoward, because the
energy spectrum is no longer flat. However several generic features are common
to the ZFL, point-particle and full NR calculations. Fig.~\ref{fig:spceNUMR}
illustrates our point.  In the left panel we show the energy spectra (rescaled
by the total ADM mass) for NR simulations of equal-mass BH collisions with
$E=1.5$ and varying impact parameter . In the right panel we show spectra for
point particles falling into a Schwarzschild BH of mass $M$ with energy
$E=1.5$ and different impact parameters. The angular momentum of the particle
normalized by the critical angular momentum (which for $E=1.5$ is $L_{\rm
  crit} \simeq 6.35M$) is indicated in the legend.  Strictly speaking, these
two plots can only be compared from a qualitative point of view. In the
equal-mass case the final BH is spinning (in fact, for large impact parameters
it is rapidly spinning \cite{Sperhake:2009jz}), whereas our point-particle
calculation considers nonspinning holes. Nevertheless, the spectra show
qualitative agreement. In particular, the ZFL is independent of the impact
parameter for both equal-mass and extreme-mass ratio collisions. For point
particles falling into Schwarzschild BHs the cutoff frequency does not depend
on $m$, but early calculations of nonrelativistic equatorial infalls into Kerr
BHs \cite{Kojima:1983ua,Kojima:1984pz,Kojima:1984cj} hint that this degeneracy
in the cutoff frequency should be lifted when one considers the rotating case.

For small, but finite frequencies, we find that the spectrum of positive-$m$
(negative-$m$) modes has positive (negative) slope as $\omega\to 0$. The
ZFL-inspired calculation for generic impact parameter of section
\ref{zflgrazing} taught us that while the ZFL itself is a robust feature, the
finite-frequency behavior is not, and it strongly depends on the modeling of
constraining forces. Such arbitrariness is absent in the point-particle
calculation.

A fit to NR results for $E=1.5$ and $l=m=2$ yields
\be
\frac{1}{M^2}\frac{dE}{d\omega}=
\frac{1}{M^2}\frac{dE}{d\omega}\Big| _{\omega =0}+
0.15\frac{L_z}{L_{\rm crit}} M\omega\,.
\ee
In the point-particle limit, a fit of the $l=m=2$ spectra with $E=1.5$ yields
\be
\frac{1}{(\mu E)^2}\frac{dE}{d\omega}=
\frac{1}{(\mu E)^2}\frac{dE}{d\omega}\Big| _{\omega =0}+
0.65\frac{L_z}{L_{\rm crit}} M\omega\,.
\ee

%%%%%%%%%%%%%%%%%%%%%%%%%%%%%%%%%%%%%%%%%%%%%%%%%%%%%%%%%%%%%%%%%%%%%%%%%%%%%%
\section{Outlook\label{conclusions}}
%%%%%%%%%%%%%%%%%%%%%%%%%%%%%%%%%%%%%%%%%%%%%%%%%%%%%%%%%%%%%%%%%%%%%%%%%%%%%%

In this paper we have used a combination of ZFL calculations and perturbative
techniques to study some of the main features emerging from ongoing NR
simulations of ultrarelativistic BH collisions. Here we wish to point out
possible extensions of our analysis.

We are currently working on a perturbative analysis of ultrarelativistic
infalls with generic impact parameter into Kerr BHs (see
\cite{Kojima:1983ua,Kojima:1984pz,Kojima:1984cj} for studies of infalls from
rest).  Another obvious generalization would be to extend ZFL and
point-particle calculations to higher-dimensional spacetimes. A preliminary
investigation of head-on, ultrarelativistic infalls into $D$-dimensional
(Schwarzschild-Tangherlini) BHs can be found in Ref.~\cite{Berti:2003si}. It
would be interesting to extend that study to plunges with generic energy and
impact parameter, and eventually also to rotating (Myers-Perry) BHs.

As mentioned in the introduction, NR simulations have provided evidence of the
existence of two critical thresholds, depending on the impact parameter $b$ of
the collision: for $b < b^*$ (where $b^*$ is the {\em threshold of immediate
  merger}) the BHs merge within the first encounter; for $b^* < b < b_{\rm
  scat}$ (where $b_{\rm scat}$ is the {\em scattering threshold}) the binary
does not merge immediately, but sufficient energy is radiated to put the
binary into a bound state that {\it eventually} results in a merger; finally,
for $b > b_{\rm scat}$ the BHs scatter producing bremsstrahlung radiation
\cite{Sperhake:2009jz}. The emergence of these two different thresholds $b_*$
and $b_{\rm scat}$ can be explained in terms of a gravitational
radiation-induced splitting of the scattering threshold. This splitting can be
described in terms of the so-called Melnikov function, which is well-known
from the theory of dynamical systems \cite{Arnold}. A detailed analysis of
this problem will be the topic of a future publication.

In this paper we have not discussed the radiation emitted by ultrarelativistic
encounters leading to scattering, rather than BH formation. Studies of
gravitational bremsstrahlung started in the sixties with the work by Peters
\cite{Peters:1970mx}, and they employed several approximation schemes. Turner
and Will studied the gravitational radiation emitted as a result of the
scattering in a low-velocity (post-Newtonian) approximation
\cite{1976ApJ...210..764W,Turner:1977,Turner:1978}. D'Eath and collaborators
developed an elegant approximation to estimate the radiation in the
ultrarelativistic case \cite{1978GReGr...9..999C,D'Eath:1976ri,D'Eath:1992hb,
  D'Eath:1992hd,D'Eath:1992qu,Segalis:2001ns}. Matzner and Nutku developed a
``method of virtual quanta'' analogous to the Weizs\"acker-Williams method of
virtual quanta in electromagnetism, which is restricted to high-velocity,
extreme-mass ratio encounters \cite{Matzner:1974rd}. The most complete study
of bremsstrahlung is perhaps a series of papers by Kovacs, Thorne and Crowley,
who computed the radiation from compact objects of arbitrary mass ratio flying
past each other at arbitrary velocities in the limit of large impact
parameter
\cite{1975ApJ...200..245T,1977ApJ...215..624C,Kovacs:1977uw,Kovacs:1978eu}.
We refer the reader to section VI of Ref.~\cite{Kovacs:1978eu} for a
comprehensive review of methods of computing gravitational bremsstrahlung
radiation.

In the eighties, various groups worked out the gravitational radiation from
particles scattering around BHs at nonrelativistic and relativistic velocities
\cite{Oohara:1984ck,Oohara:1984iu} (see also \cite{Nakamura:1987zz} for a
review). It would be interesting to repeat their perturbative calculations
paying special attention to ultrarelativistic scattering and to the
near-critical behavior.  Moreover, a comparison of bremsstrahlung
radiation as computed in NR against all these different approximations would
be a very interesting topic for future work.

Yet another semianalytical approach that we have not considered in this paper,
but that could certainly prove useful as a diagnostic of NR codes in $D$
dimensions, is the close-limit approximation (see e.g. \cite{Yoshino:2006kc}).
A detailed quantitative understanding of gravitational radiation from
$D$-dimensional BH collisions will ultimately rely on the development of NR in
higher dimensional spacetimes. Various groups are making rapid progress in
this direction
\cite{Yoshino:2009xp,Nakao:2009dc,Shibata:2009ad,Zilhao:2010sr}.

%%%%%%%%%%%%%%%%%%%%%%%%%%%%%%%%%%%%%%%%%%%%%%%%%%%%%%%%%%%%%%%%%%%%%%%%%%%%%
\section*{Acknowledgments}
%%%%%%%%%%%%%%%%%%%%%%%%%%%%%%%%%%%%%%%%%%%%%%%%%%%%%%%%%%%%%%%%%%%%%%%%%%%%%%

This work was supported by NSF grants PHY-0745779, PHY-0601459, PHY-0652995,
PHY-090003, PHY-0900735, by the Alfred P. Sloan Foundation and by the Sherman
Fairchild foundation to Caltech. It was partially funded by FCT - Portugal
through projects PTDC/FIS/64175/2006, PTDC/FIS/098025/2008,
PTDC/FIS/098032/2008, PTDC/CTE-AST/098034/2008 and
CERN/FP/109290/2009. V.C. is supported by a ``Ci\^encia 2007'' research
contract and by Funda\c c\~ao Calouste Gulbenkian through a short-term
scholarship. Computations were performed at TeraGrid in Texas, Magerit in
Madrid, the Woodhen cluster at Princeton University and HLRB2 Garching. The
authors thankfully acknowledge the computer resources, technical expertise and
assistance provided by the Barcelona Supercomputing Center - Centro Nacional
de Supercomputación.

%%%%%%%%%%%%%%%%%%%%%%%%%%%%%%%%%%%%%%%%%%%%%%%%%%%%%%%%%%%%%%%%%%%%%%%%%%%%%%%%
\appendix

%%%%%%%%%%%%%%%%%%%%%%%%%%%%%%%%%%%%%%%%%%%%%%%%%%%%%%%%%%%%%%%%%%%%%%%%%%%%%%%%
\section{The Sasaki-Nakamura formalism\label{sec:SNformalism}}
%%%%%%%%%%%%%%%%%%%%%%%%%%%%%%%%%%%%%%%%%%%%%%%%%%%%%%%%%%%%%%%%%%%%%%%%%%%%%%%%

The gravitational radiation generated by point particles in BH backgrounds is
best described in the Sasaki-Nakamura formalism
\cite{SasakiNakamura,Sasaki:1981kj,Sasaki:1981sx}. Here we summarize the
equations describing the general infall of particles of rest mass $\mu$ with
arbitrary energy (per unit rest mass) $E$.  We consider the metric of a
nonrotating BH in Schwarzschild coordinates:
\be
ds^2=-f(r)dt^2+f(r)^{-1}dr^2+r^2\left(d\theta^2+\sin^2\theta d\phi^2\right)\,,
\ee
where $f(r)\equiv 1-2M/r$ and $M$ is the BH mass.
In the Teukolsky formalism the
perturbation equations can be reduced to a second-order differential
equation for the Newman-Penrose scalar $\psi_4$ 
with a source term $T_T$. We can expand $\psi_4$ as
\be
\psi_4(t,r,\Omega)=r^{-4}\int^\infty_{-\infty}d\omega\sum_{lm}R_{lm}(r)~_{-2}Y_{lm}(\Omega)e^{-i\omega
  t}, \label{psi4expansion}\ee
and similarly for $T_T$. We denote by ${_s}Y_{lm}(\theta)$ the spin-$s$ weighted spherical harmonic, which can be expressed in terms of the well-known
(scalar) spherical harmonics \cite{Goldberg:1966uu}.

For a point particle falling into a BH along a geodesic the coordinates can be
parametrized in terms of (say) the radial location of the particle:
$(t,r,\Omega)=(T(R),R,\Omega(R))$. The source term in Teukolsky's equation is
directly related to the energy-momentum tensor $T^{\mu\nu}$ of the test
particle of mass $\mu$:
\be T^{\mu \nu}=\frac{\mu}{r^2|\dot{r}|}\delta(t-T(R))\,
\delta^2(\Omega-\Omega(R))\frac{dx^{\mu}}{d\tau}
\frac{dx^{\mu}}{d\tau}\, , \ee where an overdot denotes $d/d\tau$.
In particular,
\be T_T\equiv 4\int\, d\Omega\, dt \rho^{-6}\left
(B_{2~lm}'+B_{2~lm}^{*'}\right ) e^{-im\phi}{_{-2}}Y_{lm}(\theta)e^{i\omega
t}\,, \ee
with
\beq B_{2~lm}'&=&-\frac{1}{2}\rho^9{\cal L}_{-1}\lbrack \rho^{-4}  {\cal
L}_{0} \left (\rho^{-3}T_{nn}\right )\rbrack
\nonumber \\
&+& \frac{1}{2\sqrt{2}}\rho^9\Delta^2{\cal L}_{-1} \lbrack
\rho^{-2}J_{+}
\left (\Delta^{-1}\rho^{-4}T_{n\bar{m}}\right )\rbrack\,,\\
B_{2~lm}^{*'}&=&-\frac{1}{4}\rho^9\Delta^2 J_{+}\lbrack \rho^{-4} J_{+}
\left (\rho^{-1}T_{\bar{m}\bar{m}}\right )\rbrack
\nonumber \\
&+&\frac{1}{2\sqrt{2}}\rho^9\Delta^2 J_{+} \lbrack \rho^{-2}{\cal
L}_{-1} \left (\Delta^{-1}\rho^{-4}T_{n\bar{m}}\right )\rbrack\,.
\eeq

Here and below we generally omit the subscripts $(l,m)$ to simplify the
notation.
We also define $\Delta\equiv r^2f(r)=r^2-2Mr,\,\rho=-1/r$ and introduce the
differential operators
\beq
{\cal L}_{s}&=&\partial_{\theta}+\frac{m}{\sin\theta}+s\cot\theta\,,\\
J_{\pm}&=&\partial_{r}\pm i\frac{\omega}{f}\,. \eeq
The quantities $T_{nn} \equiv T_{\mu \nu} n^{\mu} n^{\nu}$,
$T_{n\bar{m}} \equiv T_{\mu \nu} n^{\mu} \bar{m}^{\mu}$, and
$T_{\bar{m}\bar{m}} \equiv T_{\mu \nu} \bar{m}^{\mu} \bar{m}^{\nu}$
%$T_{nn}\equiv T^{\mu\nu}n_{\mu}n_{\nu},\,T_{n\bar{m}}\equiv T^{\mu\nu}n_{\mu}\bar{m}_{\nu},\,T_{\bar{m}\bar{m}}\equiv %T^{\mu\nu}\bar{m}_{\mu}\bar{m}_{\nu}$
are contractions of $T^{\mu\nu}$ with the Kinnersley tetrad ``legs''
\beq
n_{\mu}&=&-\frac{1}{2}\left (f(r),1,0,0 \right )\,,\\
\bar{m}_{\mu}&=&-\frac{1}{\sqrt{2}\,r}\left (0,0,-r^2,ir^2\sin\theta
\right )\, , \eeq 
and an overbar denotes complex conjugation. The explicit expressions are
\beq T_{nn}&=& \frac{\mu}{4 r^2}\frac{1}{|\dot{r}|}\left (
f(r)\,\dot{t}+\dot{r}\right )^2\,
\delta(t-T(R))\,\delta^2(\Omega-\Omega(R))\,,\nonumber\\
T_{n\bar{m}}&=&
\frac{\mu}{2r^2}\frac{1}{|\dot{r}|}\frac{ir\,\dot{\phi}}{\sqrt{2}}
\left (f(r)\,\dot{t}+\dot{r}\right )\,\delta(t-T(R))\,
\delta^2(\Omega-\Omega(R))\,,\nonumber\\
T_{\bar{m}\bar{m}}&=&
-\frac{\mu}{2}\frac{\dot{\phi}^2}{|\dot{r}|}\,\delta(t-T(R))\,
\delta^2(\Omega-\Omega(R))\,.\nonumber \eeq
It is convenient to move the delta functions out of the operators by
performing an integration by parts. To do that, note that given two functions
$A$ and $B$
\beq \int d\Omega A {\cal L}_{s}B&=& -\int d\Omega B \left
(\partial_{\theta}-\frac{m}{\sin \theta}-(s-1)
\cot\theta\right )A\,,\nonumber \\
&\equiv& -B{\cal L}^{+}_{-(s-1)}A\,, \eeq
where we assumed that the integrals exist and we defined
\be {\cal
L}^{+}_{s}=\partial_{\theta}-\frac{m}{\sin\theta}+s\cot\theta\,. \ee
From the general properties of spin-weighted spherical harmonics
\cite{Goldberg:1966uu} we get
\beq {\cal L}^{+}_{1}{\cal L}^{+}_{2}\,\,{_{-2}}Y_{lm}
&=&\sqrt{\lambda(\lambda+2)}\,\,{_{0}}Y_{lm}\,\\
{\cal
L}^{+}_{2}\,\,{_{-2}}Y_{lm}&=&-\sqrt{(l+2)(l-1)}\,\,{_{-1}}Y_{lm}\,, \eeq
where $\lambda \equiv (l-1)(l+2)$. 
Introducing the operator $L_{+}\equiv
\frac{d}{dr_*}+i\omega$, where the tortoise coordinate $r_*$ is defined by
$\frac{dr_*}{dr}=\frac{r^2}{\Delta}$, we can write
$T_T=T_{1}+T_{2}+T_{3}$, with
\begin{widetext}
\beq T_{1}&\equiv & -2\int\, d\Omega\, dt \left (\rho^3{\cal
L}_{-1}\lbrack \rho^{-4} {\cal L}_{0}\left (\rho^{-3}T_{nn}\right
)\rbrack \right ) e^{i\omega \,t-im\phi}{_{-2}}Y_{lm}(\theta)
= -\frac{\mu\,r^2}{2}\frac{e^{-im\phi+i\omega\,t}}{|\dot{r}|} \left
( f(r)\,\dot{t}+\dot{r}\right )^2{\cal L}^{+}_{1} \,{\cal L}^{+}_{2}
{_{-2}}Y_{lm}(\theta)\,,
\nonumber\\
&=& -\frac{\mu}{2r^2}\Delta^2\left (V'\right )^2\dot{r}
\sqrt{\lambda(\lambda+2)}
e^{i\omega T-im\phi(R)}\,{_{0}}\bar{Y}_{lm}(\theta_0)\,,\\
T_{2}&\equiv & 4\int\, d\Omega\,\frac{ \Delta^2 \rho^3}{2\sqrt{2}}
e^{i\omega \,t-im\phi}{_{-2}}Y_{lm}(\theta)\left [{\cal L}_{-1}
\left (\rho^{-2}J_{+}(\Delta^{-1}\rho^{-4}T_{n\bar{m}})\right )+
J_{+}\left (\rho^{-2}{\cal
L}_{-1}(\Delta^{-1}\rho^{-4}T_{n\bar{m}})\right ) \right ]\,
\nn\\
&=&-i \mu \sqrt{\lambda}\,\Delta L_{+}\left (r^2\dot{\phi}V'\,
e^{i\omega T-im\phi}\right ) \,\,{_{-1}}\bar{Y}_{lm}(\theta_0)\,,
\\
T_{3}&\equiv &-\int\, d\Omega\,e^{i\omega \,t-im\phi}
{_{-2}}Y_{lm}(\theta)\Delta^2\rho^{3} J_{+} \left
(\rho^{-4}J_+(\rho^{-1}T_{\bar{m}\bar{m}})\right )\,
=\frac{\Delta}{r}L_{+}\left (\frac{r^6}{\Delta}L_{+} \left(\frac{\mu
r}{2\dot{r}}\dot{\phi}^2 e^{i\omega T-im\phi}\right )\right )
\,\,{_{-2}}\bar{Y}_{lm}(\theta_0)\,. \eeq
\end{widetext}
Here we introduced the advanced coordinate $V\equiv t+r_*$ and we denoted
radial derivatives with a prime (so $V^\prime=dV/dr$).

In the Sasaki-Nakamura (SN) formalism
\cite{SasakiNakamura,Sasaki:1981kj,Sasaki:1981sx} one introduces a new
wave function $X_{lm}(r)$ related to the radial Teukolsky function of Eq.
(\ref{psi4expansion}) via
\be R_{lm}=\Delta L_+\left(\frac{r^2}{\Delta}L_+\left(rX_{lm}\right)\right)\,.
\ee
The function $X_{lm}$ satisfies the differential equation (\ref{SNeq}), where the source term $S_{lm}$ is related to the Teukolsky source $T_T$ by the
relation
\be \Delta
L_+\left(\frac{r^2}{\Delta}\right)L_+\left(\frac{r^5 S_{lm}}{\Delta}\right)=-T_T\,. \ee
It is convenient to define a quantity
\be 
W_{lm}=\frac{r^5}{\Delta} S_{lm}e^{i\omega r^*}\,,\label{Wsource}
\ee
which satisfies
\be \frac{d^2W_{lm}}{dr^2}=-\frac{r^2}{\Delta^2} T_T e^{i\omega r_*}\,.
\ee
By using the equality
\be e^{i\omega r_*}\, L_{+}\left (X_{lm}(r)e^{i\omega T}\right )=
\partial_{r_*}\left (X_{lm}(r)e^{i\omega V}\right )\,,
\ee
we can write $W_{lm}=W_1+W_2+W_3$, where
\beq \nn \frac{d^2W_1}{dr^2}&=&
\sqrt{\lambda(\lambda+2)}\frac{\mu}{2}\left (V'\right )^2
\dot{r}
e^{i\omega V-im\phi} 
\,\,{_{0}}\bar{Y}_{lm}(\theta_0)\,,\\
\frac{d^2W_2}{dr^2}&=& i\mu\sqrt{\lambda}\partial_r \left (V'
\,r^2\dot{\phi}
e^{i\omega V-im\phi}
\right ) \,
{_{-1}}\bar{Y}_{lm}(\theta_0)\,,
\label{Weqns}
\\
\nn \frac{d^2W_3}{dr^2}&=& -\frac{\mu}{r}\partial_r \left
(r^4\partial_r\left (\frac{r}{2}\dot{r}^{-1} \dot{\phi}^2 
e^{i\omega V-im\phi}
\right )\right ) \,\, {_{-2}}\bar{Y}_{lm}(\theta_0)\,. \eeq
Given any function $F(r)$ we have $F(r) V' e^{i\omega V}=1/(i\omega)[(F
  e^{i\omega V})'-F'e^{i\omega V}]$. If we use this identity to integrate by
parts, rearrange terms and set $\dot r=-\gamma$, these equations reduce to
Eqs.~(A5)--(A8) in Ref.~\cite{Oohara:1984bw}.

%%%%%%%%%%%%%%%%%%%%%%%%%%%%%%%%%%%%%%%%%%%%%%%%%%%%%%%%%%%%%%%%%%%%%%%%%%%%%%%%
\subsection{Calculation of the Wronskian}
%%%%%%%%%%%%%%%%%%%%%%%%%%%%%%%%%%%%%%%%%%%%%%%%%%%%%%%%%%%%%%%%%%%%%%%%%%%%%%%%

From a numerical perspective, the integration of Eq.~(\ref{SNeq}) is performed
via a standard Green's function solution. For improved accuracy, in the
numerical integrations we use asymptotic expansions of the wave functions near
the horizon and near infinity.  Close to $r=r_+=2M$ we set
$y=\frac{(r-r_+)}{r_+}$, make the ansatz
\begin{equation}
X_{lm}(r)=e^{-i\omega r^*}\left[1+\sum_{k=1}^\infty a_k y^k\right],
\end{equation}
and substitute into the differential equation (\ref{SNeq}).
%
%\begin{equation} f^2 X''(r)+f'f
%X'(r)+\left[\omega^2
%-f\left(\frac{\lambda+2}{r^2}+\frac{6}{r^3}\right)\right]=0\,.
%\end{equation}
%
Noting that $f=y-y^2+{\cal O}(y^3)$ we determine the leading-order and
next-to-leading-order coefficients to be
\beq
a_1&=&\frac{\lambda -1}{1-4 i M\omega}\,,\\
a_2&=&\frac{3+(\lambda-1+8iM\omega)a_1}{4(1-2iM\omega)}\,.\nn \eeq
In the limit $r\to \infty$ we have
\be X_{lm}\sim B^{\rm in}_{lm} e^{i\omega r^*} \left[1+\sum_{k=1}^\infty
\frac{c_k}{r^k}\right]+ A^{\rm in}_{lm} e^{-i\omega r^*}
\left[1+\sum_{k=1}^\infty \frac{d_k}{r^k}\right]. \ee
The lowest-order coefficients in the series are found to be
\beq c_1&=&\frac{i(\lambda+2)}{2\omega}\,,\,
c_2=-\frac{\lambda(\lambda+2)+12 iM\omega}{8\omega^2}\,,\\
d_1&=&-c_1\,,\, d_2=\frac{12
iM\omega-\lambda(\lambda+2)}{8\omega^2}\,. \eeq
One can invert these relations to get
\begin{widetext}
\beq A^{\rm in}_{lm}&=&-r\,\frac{\left
[(2M-r)(2c_2+c_1r)+i\,r^2(c_2+r(c_1+r))\omega \right
]X_{lm}(r)-r\,(r-2M)[c_2+r(c_1+r)]X'_{lm}(r)} {(2M-r)
\left[c_1d_2+2d_2r+(d_1-c_1)r^2-c_2(d_1+2r)\right]
+2r[c_2+r(c_1+r)][d_2+r(d_1+r)]\omega}\,e^{i\omega\,r_{*}}\,,
\label{wronsk:Ain}
\\
\nn B^{\rm in}_{lm}&=&i\,r\,
\frac{\left [i\,(2M-r)(2d_2+d_1r)+r^2(d_2+r(d_1+r))\omega\right]
X_{lm}(r)-i\,r\,(r-2M)[d_2+r(d_1+r)]X'_{lm}(r)} {(2M-r)\left
[-c_1d_2-2d_2r+(c_1-d_1)r^2+c_2(d_1+2r)\right]
+2i\,r[c_2+r(c_1+r)][d_2+r(d_1+r)]\omega}\,e^{-i\omega\,r_{*}}\,,
\eeq
\end{widetext}
where a prime stands for a derivative with respect to $r$ and in our numerics
all quantites are evaluated at large but finite $r$ (typically at
$r=r_{\infty}=r_{\infty}^{(0)}/\omega$, with $r_{\infty}^{(0)} = 4\times
10^4$).  The Wronskian appearing in the Green's function solution of the
inhomogeneous equation can be evaluated from the relation $W=2i\omega A^{\rm
  in}_{lm}$.
%

%%%%%%%%%%%%%%%%%%%%%%%%%%%%%%%%%%%%%%%%%%%%%%%%%%%%%%%%%%%%%%%%%%%%%%%%%%%%%%%%
\section{Multipolar decomposition of ZFL spectra\label{app:multipoles}}
%%%%%%%%%%%%%%%%%%%%%%%%%%%%%%%%%%%%%%%%%%%%%%%%%%%%%%%%%%%%%%%%%%%%%%%%%%%%%%%%

Here we discuss the extraction of multipolar components from the ZFL
calculation. For consistency with the conventions we used in NR simulations
\cite{Sperhake:2008ga,Sperhake:2009jz} and in the point-particle infalls
discussed in this paper, we assume that the polarization states have an
angular dependence that can be decomposed in spin-weighted spherical
harmonics: $h_+-ih_\times \sim \sum h_{lm}{_{-2}}Y_{lm}$. Then (using the
completeness of spin-weighted spherical harmonics) we can extract the
multipolar content of the energy spectrum by equating
\be \frac{d^2E}{d\omega d\Omega}=\left(\sum_{lm}\,\sqrt{\frac{dE_{lm}}{d\omega}}\,
{_{-2}}Y_{lm}\right )^2\,,
\ee
where $dE_{lm}/d\omega$ are (as yet undetermined) functions of $\omega$. From
the orthonormality of spin-weighted spherical harmonics we find
\be 
\sqrt{\frac{dE_{lm}}{d\omega}}=\int d\Omega \sqrt{\frac{d^2E}{d\omega
    d\Omega}}{_{-2}}Y_{lm}\,.
\ee
We now determine analytically the multipolar content of the radiation in the
simple case of ultrarelativistic head-on collisions.

%%%%%%%%%%%%%%%%%%%%%%%%%%%%%%%%%%%%%%%%%%%%%%%%%%%%%%%%%%%%%%%%%%%%%%%%%%%%%%%%%
\subsection{Equal mass collisions}
%%%%%%%%%%%%%%%%%%%%%%%%%%%%%%%%%%%%%%%%%%%%%%%%%%%%%%%%%%%%%%%%%%%%%%%%%%%%%%%%%

If we assume that the collision occurs along the $z$-axis, the radiation will
only contain $m=0$ modes.  For equal-mass, ultrarelativistic head-on
collisions Eq.~(\ref{b0}) implies
\beq \frac{dE_{l0}}{d\omega}&=&\frac{4 E^2 M^2}{\pi}
\frac{(2l+1)(l-2)!}{(l+2)!}\,,\quad l \,\,\quad {\rm even}\nonumber\\
&=&0\,,\hskip3.8cm l \quad \,{\rm odd} \,.\label{energyspectrumequal}
\eeq
Odd multipoles do not contribute, as required by equatorial symmetry. Summing
over even multipoles we get $dE/d\omega=E^2 M^2/\pi$, in agreement with
Smarr's Eq.~(2.20) when $v\to c$ \cite{Smarr:1977fy}.  This expression assumes
that the collision occurs along the $z-$axis, but the multipolar components in
a general (rotated) frame can be found following the procedure discussed in
Ref.~\cite{Gualtieri:2008ux}.

%%%%%%%%%%%%%%%%%%%%%%%%%%%%%%%%%%%%%%%%%%%%%%%%%%%%%%%%%%%%%%%%%%%%%%%%%%%%%%%%
\subsection{Extreme-mass ratio collisions}
%%%%%%%%%%%%%%%%%%%%%%%%%%%%%%%%%%%%%%%%%%%%%%%%%%%%%%%%%%%%%%%%%%%%%%%%%%%%%%%%

From Eq.~(\ref{mMb}), for $\mu \equiv M_1\ll M_2 \equiv M$ we get
\be \frac{dE_{l0}}{d\omega}=\frac{4 E^2
\mu^2}{\pi}\frac{(2l+1)(l-2)!}{(l+2)!}\,. \label{energyspectrumunequal}\ee
Summing over multipoles we get $dE/d\omega=4/(3\pi)E^2 \mu^2$, in
agreement with Smarr's Eq.~(2.18) when $v\to c$ \cite{Smarr:1977fy}.

We recall once more that Eq.~(\ref{energyspectrumunequal}) is valid in a frame
where the collision occurs along the $z$--axis, so only $m=0$ modes are
present. The transformation to a general (rotated) frame is explained in
Ref.~\cite{Gualtieri:2008ux}.  In the coordinate system used to compute the
radiation from point particles falling into a Schwarzschild BH
(section~\ref{ppart}), and focusing on the dominant ($l=2$) components, we
get:
\beq \frac{dE_{20}}{d\omega}&=&\frac{5 E^2
\mu^2}{24\pi}=0.0663146 E^2
\mu^2\,,\\
\frac{dE_{22}}{d\omega}&=&\frac{5 E^2
\mu^2}{16\pi}=0.0994718 E^2
\mu^2\,.
\eeq
This is in excellent agreement with the point-particle results listed in
Table~\ref{tab:zfl}.

%%%%%%%%%%%%%%%%%%%%%%%%%%%%%%%%%%%%%%%%%%%%%%%%%%%%%%%%%%%%%%%%%%%%%%%%%%%%%%%%
%\bibliography{BibPP}

\end{document}